\documentclass{aastex631-1}

\usepackage{placeins}
\usepackage{morefloats}
\usepackage{graphicx}

\shorttitle{LISM Cloud Temperatures}
\shortauthors{Linsky and Redfield}

\begin{document}

\title{Inhomogeneity within Local Interstellar Clouds\footnote{Based on
observations made with the NASA/ESA
Hubble Space Telescope obtained from the Data Archive at the Space
Telescope Science Institute, which is operated by the Association of
Universities for Research in Astronomy, Inc., under NASA contract NAS
AR-09525.01A.}}

\author[0000-0003-4446-3181]{Jeffrey L. Linsky}
\affiliation{JILA, University of Colorado and NIST, Boulder, CO 
80309-0440, USA}

\author{Seth Redfield}
\affiliation{Astronomy Department and Van Vleck Observatory,
  Wesleyan University, Middletown, CT 06459-0123, USA}
 
\author{Diana Ryder}
\affiliation{Department of Astrophysical and Planetary Sciences, University of Colorado, Boulder, CO, USA} 

\author{Adina Chasan-Taber}
\affiliation{Astronomy Department and Van Vleck Observatory,
  Wesleyan University, Middletown, CT 06459-0123, USA}
 
\correspondingauthor{Jeffrey L. Linsky}
\email{jlinsky@jila.colorado.edu}




\begin{abstract}
Analysis of interstellar absorption lines observed in high-resolution {\em HST} spectra of nearby stars provides temperatures, turbulent velocities, and kinetic properties of warm interstellar clouds. Previous studies identified 15 warm partially ionized clouds within about 10~pc of the Sun and measured their mean thermal and kinematic properties. A new analysis of 100 interstellar velocity components reveals a wide range of temperatures and turbulent velocities within the Local Interstellar Cloud (LIC) and other nearby clouds. These variations appear to be random with Gaussian distributions. We find no trends of these properties with stellar distance, angle from the Galactic Center, the main source of EUV radiation (the star   $\epsilon$~CMa), the center of the LIC, or the direction of inflowing interstellar matter into the heliosphere. The spatial scale for temperature variations in the LIC is likely smaller than 5,100~au, a distance that the Sun will traverse in 1,000 years. Essentially all velocity components align with known warm clouds. We find that within 4~pc of the Sun, space is completely filled with partially ionized clouds, but at larger distances space is only partially filled with partially ionized clouds, indicating that fully ionized inter-cloud gas fills the voids. We find that the neutral hydrogen number density in the LIC and likely other warm clouds in the CLIC is about 0.10~cm$^{-3}$ rather than the 0.20~cm$^{-3}$ density that may be representative of only the immediate environment of the LIC.  
The 3,000--12,000~K temperature range for the gas is consistent with the predictions of theoretical models of the WNM and WIM, but the high degree of inhomogeneity within clouds argues against simple theoretical models. Finally, we find evidence for a shock in the sight line to the star AD~Leo.
\end{abstract}


\keywords{Interstellar medium (847), interstellar dynamics (839), Interstellar
  magnetic fields (845), Ultraviolet sources (1741)}

\section{Introduction}

    The heliosphere does not live in static isolation, but is instead encapsulated by an environment with inhomogeneous spatial properties that manifest as time variable external properties as the heliosphere traverses the interstellar medium. Total pressure balance between the outer heliosphere, hereafter called the very local interstellar medium (VLISM),  and the surrounding interstellar medium can greatly alter the size and shape of the heliosphere. For example, Zank \& Frisch (1999) and M\"uller et al. (2006)  computed heliospheric models for a wide range of interstellar pressures showing that when the heliosphere enters a cold cloud with density 100,000 times larger than the Local Interstellar Cloud (LIC), the heliosphere would shrink to the size of the inner solar system. The density and ionization within the heliosphere will respond to time variations in the interstellar density, ionization, flow vector, and magnetic field strength. In order to estimate the possible range of physical parameters in the heliosphere over time, it is essential to explore the range of interstellar environments that the heliosphere has and will encounter. Our understanding of heliospheric evolution provides the basis for modeling astrospheres and their interactions with exoplanet atmospheres. 
     
In early theoretical models of the interstellar medium \citep[e.g.][]{Field1969, Wolfire1995}, the balance between heating and cooling processes in a constant pressure environment predicts that thermal instability will drive the interstellar medium into distinct phases each characterized by a narrow range of temperatures and densities. These phases have been called the cold neutral medium (CNM) ($T<300$~K), warm neutral medium (WNM) ($T\approx 8,000$~K), warm ionized medium (WIM) ($T\approx 8,000$~K), and hot ionized medium (HIM) ($T\approx 10^6$~K). These theoretical models were constructed assuming time independent steady state populations and pressure equilibrium among the phases. A very different model was identified by simulations that include supernova and stellar wind heating and dynamics and time dependent evolution of gas without the constraint of constant pressure. In the simulations of \cite{deAvillez2005}, turbulence drives the wide range of values for local temperatures, densities, magnetic field strengths, and flows. Given these two very different models for the interstellar medium, it is important to measure the properties of local interstellar gas because proximity provides the best angular resolution, and short path lengths to nearby stars removes some of the complexity inherent in long sight lines through the interstellar medium. A  critical question is whether the local interstellar medium (LISM) is more like the first or second type of models, or perhaps contains aspects of both models.

This paper addresses questions concerning the inhomogeneity of the LISM -- the range of temperatures and turbulent velocity variations inside the Local Interstellar Cloud (LIC) and other nearby partially ionized clouds, possible trends in these inhomogeneities, possible causes for these inhomogeneities, likely variations in the heliosphere resulting from the trajectory of the heliosphere through the inhomogeneous LISM, and whether the LISM consist of warm clouds that partially or completely fill the LISM. The review paper entitled "Inhomogeneity in the local ISM and its relation to the heliosphere" \citep{Linsky2022} contains a preliminary presentation and discussion of these issues based upon a smaller data set. Previous reviews include \cite{Redfield2006} and \cite{Frisch2011}.

In Section 2 we describe our measurement techniques and previous measurements of temperatures, turbulent velocities and kinematics of nearby warm interstellar clouds. Section 3 describes the inhomogeneous properties within the LIC and other clouds, and possible trends with different parameters. In Section 4 we test the predictions of theoretical models against our data, and Section 5 summarizes our results.

\section{Measurement of Cloud Temperatures and Turbulence}

    Our knowledge of the local interstellar medium (LISM) is based primarily on high-resolution ultraviolet spectra of stars that include narrow absorption lines produced by interstellar gas in the line of sight to a more distant star. The most useful interstellar lines are transitions from the ground states of H~I (Lyman-$\alpha$, 1215.67~\AA), D~I (Lyman-$\alpha$, 1215.34~\AA), Mg~II (2796.35, 2803.53~\AA), Fe~II (2586.65, 2600.127~\AA), and other lines of O~I, C~II, and Si~II. Interstellar absorption lines of Ca~II and Na~I in the optical spectrum are generally too weak to be seen on short sight lines. 
    
The analysis of {\em HST} absorption line profiles is described in detail by \cite{Redfield2004b} and \cite{Malamut2014}, but here is a short overview. The high-resolution STIS spectra were obtained from the .x1d files in the MAST Portal. We used a suite of software written in the IDL language to fit Voigt profiles to the narrow interstellar absorption lines observed against broad emission lines formed in stellar chromospheres. We first fit the stellar emission line, typically with a self-reversal, by a sixth degree polynomial, and then simultaneously fit one or more interstellar velocity components that provide an optimal fit to the observed spectrum. For Mg~II we fit both lines in the multiplet simultaneously. We individually fit both fine structure components of the H~I and D~I Lyman-$\alpha$ lines.  The fitting procedure takes into account instrumental broadening appropriate from the time of observation. The output consists of the central wavelength and velocity of the interstellar absorption line, the line width Doppler parameter $b$, the column density of the atom or ion, the estimated hydrogen column density, and both formal errors and Monte Carlo simulation errors.  For every sightline that we have analyzed, there is at least one interstellar absorption component, but there are often two or more velocity components indicating different parcels of interstellar gas moving at different radial velocities in the sight line to the star. The average number of components is about 1.5 per sight line \citep{Malamut2014}.

The measurement of cloud temperatures $T$ and nonthermal (turbulent) broadening $\xi$ involves comparing the line widths of multiple ions of different atomic mass. The interstellar absorption line width $b$ is the sum of two components,

\begin{equation}
b^2=2kT/m + \xi^2 = 0.016629T/A + \xi^2,
\end{equation}
 where $k$ is Boltzmann's constant, $m$ is the mass of the atom or ion, and $A$ is the atomic weight of the atom or ion in atomic mass units. This equation assumes that
the observed plasma can be described by a single Maxwell-Boltzmann velocity distribution with macroscopic random flows along the line of sight through the interstellar cloud.
This simple representation need not be the case if the temperature is not uniform along the line of sight or the plasma has a microscopic supra-thermal component that could produce extended wings or Doppler shifts in the line profile. While these complex phenomena are observed in the solar corona, they have not been investigated for interstellar clouds. We therefor assume that $T$ and $\xi$ represent mean values through a cloud and proceed with the standard formula until evidence emerges that requires a more detained analysis. 

The separation of thermal from nonthermal macroscopic broadening takes advantage of the different atomic weights of the observed interstellar atoms and ions. In principle, the separation only requires line widths obtained from high resolution spectra of a low mass atom such as D~I and a high mass ion such as Mg~II or Fe~II, because the line widths of a low mass atom are primarily thermal, whereas the line widths of the high mass ion are primarily due to gas motions along the sight line. The D~I Lyman-$\alpha$ line separated from the H~I Lyman-$\alpha$ line by --0.33\AA\ is preferred for interstellar studies, because it is generally optically thin whereas the H~I line is very optically thick. The D~I line is actually a hyperfine structure doublet with components at 1215.3430 and 1215.3373~\AA. Because of the relatively weak thermal broadening of heavy ions, especially Fe~II, these ions best display the multiplicity of velocity components. However, velocity components separated by less than the resolution of the spectrograph are indistinguishable with the analysis providing a large value of non- thermal broadening. Thus high resolution spectra are essential for this work. 

Since the line widths range from $b\approx 3$ km~s$^{-1}$ for Fe~II to $b\approx 8$ km~s$^{-1}$ for D~I, high resolution spectra with an accurate absolute wavelength scale are essential. For the 1200--1700~\AA\ spectral range, the best available spectra are obtained with the E140H grating (3 km~s$^{-1}$ resolution) or E140M grating (6 km~s$^{-1}$) on the {\em HST} STIS instrument or ECH-A (3.6~km~s$^{-1}$) grating on the HRS instrument. For the 2000-3000~\AA\ spectral range the best available spectra are obtained with the 
STIS E230H (3 km~s$^{-1}$) or HRS ECH-B (3.3~km~s$^{-1}$) gratings.

\begin{figure}
\figurenum{1a}
\plottwo{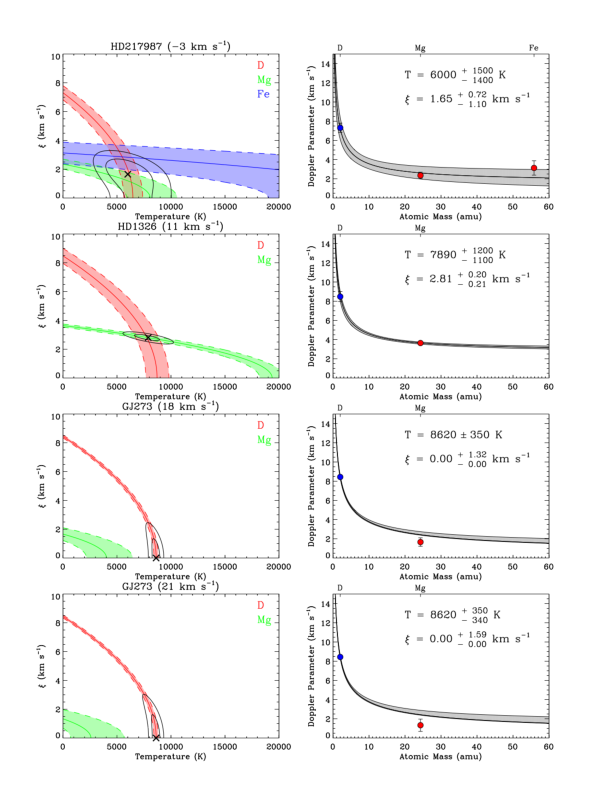}{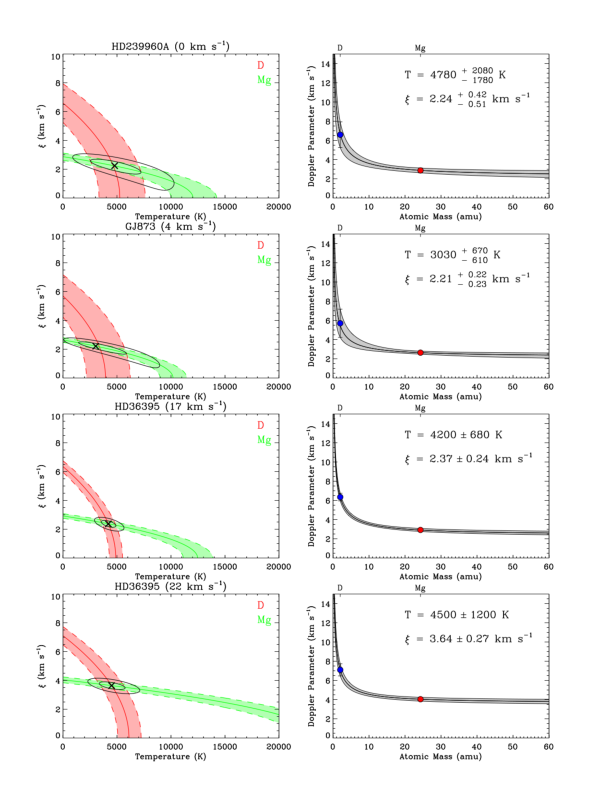}
\caption{{\bf Left:} Plots of turbulent velocity vs. temperature for different ions, best fit (cross), and $1\sigma$ and $2\sigma$ contours about the best fit. {\bf Rght:} Doppler parameters vs mass for each ion included in the analysis. The best fit parameters and $1\sigma$ contours fitting the Doppler parameters are included. \label{temp6-0}}
\end{figure}

Fig.~\ref{temp6-0} and Fig.~\ref{temp6-1}  show how the the analysis of line widths of low and high mass ions leads to a best estimate of the mean temperature and turbulent velocity in the sight lines to stars that we have measured.

\begin{figure}
\figurenum{1b}
\plottwo{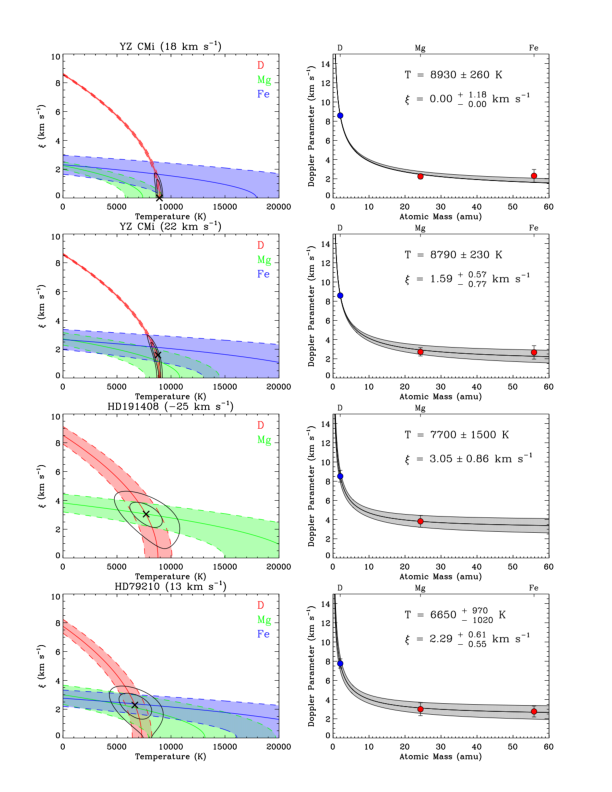}{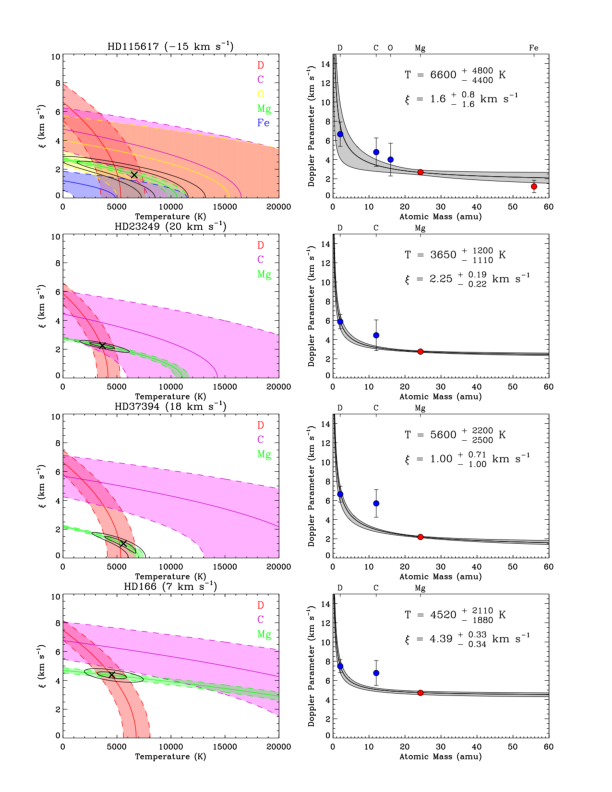}
\caption{}
\end{figure}

\begin{figure}
\figurenum{1c}
\plottwo{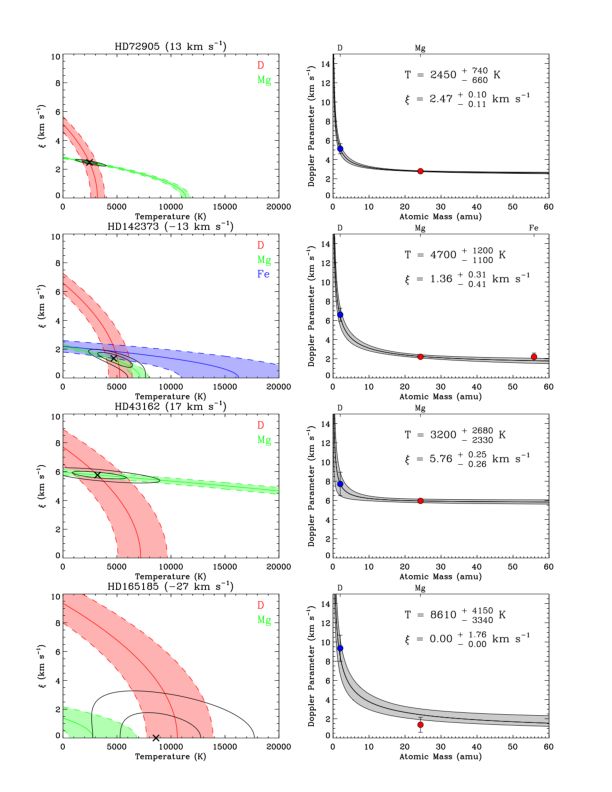}{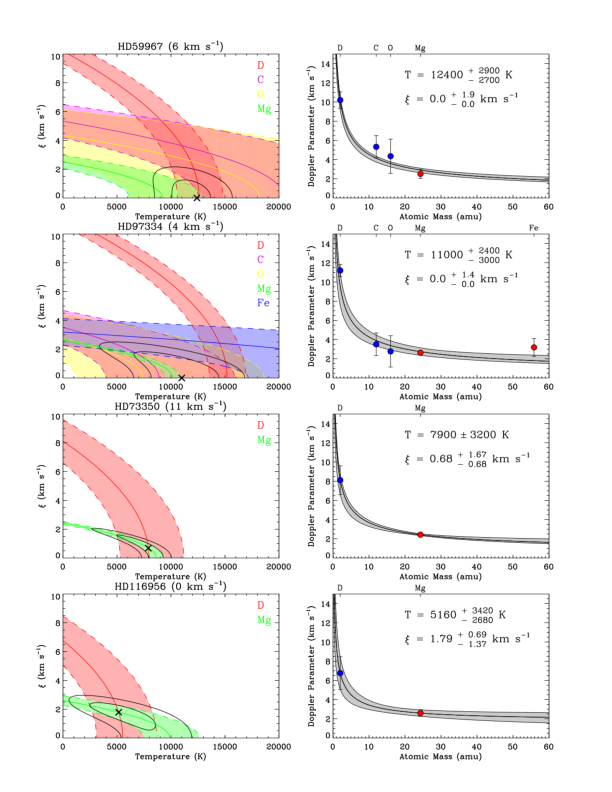}
\caption{}
\end{figure}

\begin{figure}
\figurenum{1d}
\plottwo{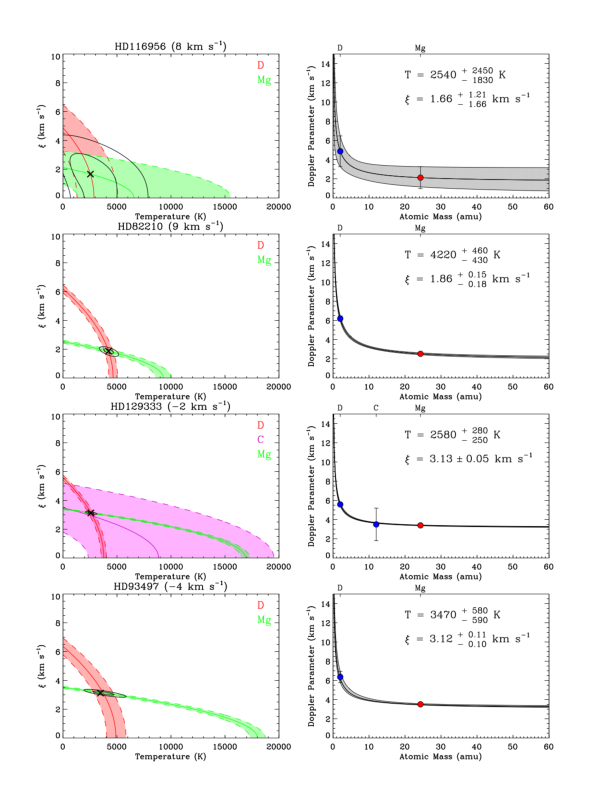}{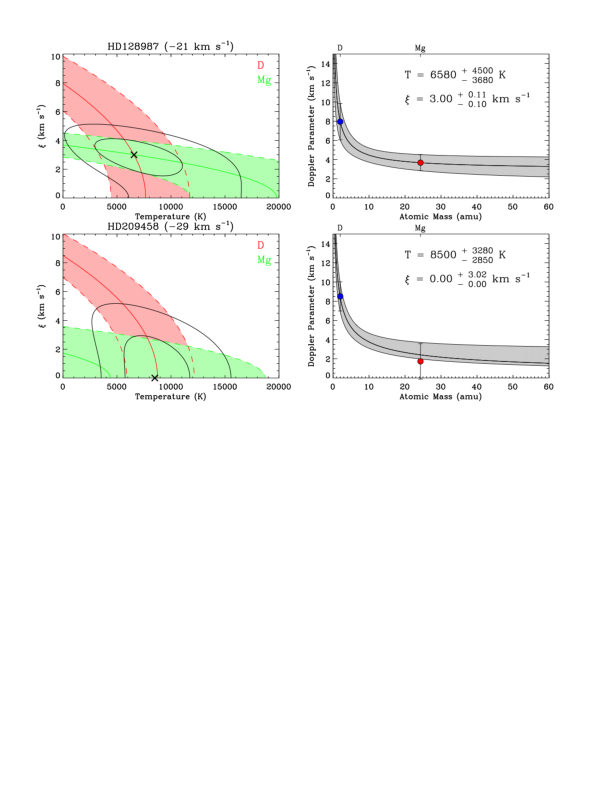}
\caption{}
\end{figure}

The measured neutral hydrogen column density plays a critical role in determining the size of interstellar clouds (see Section 3.11), but the analysis of the only available neutral hydrogen line with interstellar absorption, Lyman-$\alpha$, poses particular challenges. The line core is extremely optically thick, often $10^6$ even for short sight lines, the line has broad damping wings that require a Voigt profile for fitting, there may be a central self-reversal produced by the loss of photons near the top of the chromosphere \citep[cf.,][]{Youngblood2022}, and the narrow geocoronal emission feature located near line center must be removed before the line profile can be fit. In addition, charge exchange between hydrogen atoms inflowing from the LISM and protons produces regions of decelerated, hot, and relatively dense neutral hydrogen in what are called hydrogen walls in the outer heliosphere and astrospheres of stars similar to and cooler than the Sun. \cite{Wood2005b} described the analysis of Lyman-$\alpha$ lines that take into account the extra absorption on the red side of the interstellar absorption from the heliospheric hydrogen wall and the extra absorption mostly on the blue side from the astrospheric hydrogen wall. Since these extra absorptions can be weak and difficult to identify and fit accurately, the analysis of the hydrogen Lyman-$\alpha$ line by itself needs a further constraint. The neutral deuterium Lyman-$\alpha$ line centered at --0.33~\AA\ from the hydrogen Lyman-$\alpha$ line provides this constraint because the deuterium line optical depth is often optically thin and $N$(D~I) is too small for any significant deuterium wall absorption. The neutral hydrogen column density can be inferred from the deuterium line, $N$(H~I)=$N$(D~I)/$R$, where $R=1.5\times10^{-5}$ is the deuterium/hydrogen number ratio within 100~pc of the Sun \citep{Wood2004,Linsky2006}. Most $N$(H~I) data comes from simultaneously fitting both the hydrogen and deuterium lines, but highly constrained $N$(H~I) results can be obtained by fitting only the deuterium line.

\subsection{Previous Measurements}     

The first measurements of interstellar temperatures, turbulent velocities, and kinematics were for individual bright stars, e.g.,  Capella (Linsky et al. 1993), Sirius~A (Lallement et al. 1994), Procyon (Linsky et al. 1995), $\epsilon$~CMa (Gry et al. 1995, Gry \& Jenkins (2001), $\alpha$~Cen (Linsky \& Wood 1996), and $\beta$~CMa (Jenkins et al. 2000).
\cite{Redfield2004b} and \cite{Redfield2008} provided the first analysis of a large number of sight lines using the spectrographs on the {\em Hubble Space Telescope (HST)}. They measured temperatures, nonthermal broadening, and heliospheric velocities for 50 velocity components in the sight lines toward 29 stars located within 100~pc. Fourteen of the sight lines show only one velocity components, nine show two components, and six show three components. All of the stars show at least one interstellar velocity component. Thus interstellar absorption observed in the H~I, D~I, and Mg~II lines is ubiquitous with multiple velocity components common even for lines of sight shorter than 3.3~pc. 

\cite{Redfield2004a} found a weighted mean temperature and weighted dispersion $<T>=6900^{+2400}_{-2100}$~K. The distribution does not appear Gaussian about the mean value as there are at least 8 velocity components with temperatures well below the expected for a Gaussian distribution with this standard deviation. Additional observations are needed to better characterize the distribution of temperatures. They also found that the mean value of nonthermal broadening is $<\xi>=2.2\pm 1.03$~km~s$^{-1}$ with an apparent excess of large $\xi$ values that could result from unresolved closely space velocity components. Subsequently, \cite{Redfield2008} found the weighted mean temperature of $7,500\pm 1,300$~K for the 19 LIC stars in their sample. Table~1 compares the weighted mean temperatures and weighted dispersions about the mean temperatures obtained from past and present data sets. For comparison, we include the temperature and its error for neutral helium gas flowing into the heliosphere from the LIC \citep{Swaczyna2018}. There is excellent agreement between the mean temperatures for the LISM and LIC data sets with the temperatures measured for neutral helium flowing into the heliosphere.

\begin{table}
\caption{Temperature Measurements for Local Interstellar Gas} 
\begin{center}
\begin{tabular}{lcccc}
\hline\hline
Region & Components & Mean Temperature & Temperature Range & Reference\\ \hline
Inflow of LIC He~I gas & 1 & $7,691\pm 230$~K & -- & \cite{Swaczyna2018}\\
LIC & 19 & $7,500\pm 1,300$~K & 5,200--12,900~K & \cite{Redfield2008}\\
LIC ($0 \sigma$) &  37  & $6,509\pm 2,761$~K  &  1,830--12,900~K  & This paper\\
LIC ($2\sigma$) & 36 & $6,511\pm 2,773$~K & 2,450--12,900~K & This paper\\
Warm clouds & 50 & $6,900^{+2,400}_{-2,100}$~K & 1,000--12,600~K & \cite{Redfield2004a}\\
Warm clouds ($0\sigma$)& 100   & $6,742\pm 2,441$~K  & 1,700--12,900~K  & This paper\\
Warm clouds ($2\sigma$) & 84 & $6,838\pm 2,455$~K &  2,450--12,900~K  & This paper\\ \hline\hline
\end{tabular}
\end{center}
\end{table}

Following the \cite{Redfield2008} paper there have been two papers with additional values of $T$ and $\xi$ for other sight lines. \cite{Zachary2018} studied two sight lines each with two velocity components, and \cite{Edelman2019} studied three stars each with three velocity components. The results of these two studies are included in Table~2. 
The results of our analysis of 27 new sight lines with 31 velocity components are listed in Table~3. The addition of these 47 velocity components listed in Tables 2 and 3 to the initial list of 50 sight lines obtained by Redfield \& Linsky (2004) warrants a reexamination of the distribution of cloud temperatures and turbulent velocities.
The analysis of the new sight lines in this paper followed the approach and using the same software as described in detail by Redfield \& Linsky (2004).

\begin{table}
\caption{Temperature and Turbulent Velocity Measurements Since 2008} 
\begin{center}
\begin{tabular}{lccccccccc}
\hline\hline
HD & Name & $l$ & $b$ & d(pc) & $<v>$ & $T$ & $\xi$ & Ions Used & Ref\\ \hline 
190248 & $\delta$ Pav & 329.8 & --32.4 & 6.10 & $-17.2\pm1.5$ & $8680^{+740}_{-780}$ & $0.0^{+2.18}_{-0.0}$ & D I, C II, O I, Mg II & 1\\
190248 & $\delta$ Pav & 329.8 & --32.4 & 6.10 & $-9.23\pm0.58$ & $9310^{+10070}_{-7959}$ & $2.44^{+1.04}_{-2.44}$ & D I &1\\
192310 & GJ 785 & 15.6 & -29.4 & 8.81 & $-30.41\pm 0.57$ & $8600^{+2000}_{-1800}$ & $3.3^{+1.2}_{-1.3}$ & D I, C II, Mg II, Fe II& 2\\ 
192310 & GJ 785 & 15.6 & -29.4 & 8.81 & $-24.24\pm0.50$  & $9900^{+2200}_{-2100}$ & $2.2^{+1.1}_{-2.0}$ & D I, C II, Mg II, Fe II& 2\\
192310 & GJ 785 & 15.6 & -29.4 & 8.81 & $-18.88\pm0.57$ & $6900^{+2600}_{-2300}$ & $1.3^{+1.6}_{-1.3}$ & D I, C II, Mg II, Fe II& 2\\
HIP85665 & GJ 678.1A & 28.6 & 20.5 & 10.12 & $-31.5\pm2.4$ & $10720^{+4730}_{-3860}$ & $0.0^{+1.92}_{-0.0}$ & D I, Mg II & 1\\
HIP85665 & GJ 678.1A & 28.6 & 20.5 & 10.12 & $-23.9\pm2.4$ & $8540^{+3240}_{-2790}$ & $0.0^{+1.34}_{-0.0}$ & D I, Mg II & 1\\
9826 &   $\upsilon$~And & 132.0 & --20.7 & 13.48 & $9.1\pm 1.2$ & $10400^{+2000}_{-1900}$ & $0.00^{+2.4}_{-0.0}$ & D I, O I, Mg II, Fe II& 2\\
9826 &   $\upsilon$~And & 132.0 & --20.7 & 13.48 & $12.1\pm 1.1$ & $4000^{+2800}_{2200}$ & $1.8^{+0.8}_{-1.2}$ &  D I, O I, Mg II, Fe II& 2\\
9826 &   $\upsilon$~And & 132.0 & --20.7 & 13.48 & $16.45\pm0.88$ & $6500^{+3000}_{-2600}$ & $1.3^{+0.7}_{-1.3}$ & D I, O I, Mg II, Fe II& 2\\
206860 & NN Peg & 69.9 & --28.3 & 18.13 & $-14.68\pm0.58$ & $7100^{+2800}_{-2400}$ & $1.4^{+0.6}_{-1.4}$ & D I, Mg II, Fe II& 2\\
206860 & NN Peg & 69.9 & --28.3 & 18.13 & $-8.0\pm1.0$ & $9600^{+2500}_{-2300}$ & $2.11^{+0.54}_{-0.68}$ &  D I, Mg II, Fe II& 2\\
206860 & NN Peg & 69.9 & --28.3 & 18.13 & $-5.44\pm0.79$ & $6800^{+2700}_{-2500}$ & $0.80^{+0.98}_{-0.80}$ &  D I, Mg II, Fe II& 2\\
87901   & $\alpha$~Leo & 226.4 & 48.9 & 24.31 & $8.8\pm0.2$  & $6000^{+600}_{-600}$  & $1.78\pm0.10$  & C II, N I, O I, Mg I, Mg II  & 3\\
87901   & $\alpha$~Leo & 226.4 & 48.9 & 24.31 & $14.4\pm0.1$  & $5990^{+700}_{-700}$  & $1.85\pm0.19$  & C II, N I, O I, Mg I, Mg II  & 3\\

\hline\hline
\end{tabular}
\end{center}
References: (1) \cite{Zachary2018}; (2) \cite{Edelman2019}; (3) \cite{Gry2017}.\\
\end{table}

\clearpage
\begin{table}
\caption{New Temperature and Turbulent Velocity Measurements} 
\begin{center}
\begin{tabular}{lcccccccccc}
\hline\hline
HD & Name & $l$ & $b$ & d(pc) & $<v>$ & $T$ & $\xi$ & Ions Used & Ref & cloud \\ \hline
217987 & GJ 887   & 5.1 & --66.0 & 3.29 & $-2.74\pm0.21$ & $6000^{+1500}_{-1400}$ & $1.65^{+0.72}_{-1.10}$ & D I, Mg II & 1,2 & LIC \\
1326  & GJ 15A   & 116.7 & --18.4 & 3.56 & $10.91\pm0.32$ & $7800^{+1200}_{-1100}$ & $2.81^{+0.20}_{-0.21}$ & D I, Mg II & 1, 2 & (Hya)\\
  & GJ 273   & 212.3 & 10.4 & 3.80 & $18.28\pm0.92$ & $8620\pm350$ & $0.0^{+1.32}_{-0.0}$ & HI, Mg II & 1,2 & LIC \\
  & GJ 273   & 212.3 & 10.4 & 3.80 & $21.38\pm0.40$ & $8620^{+350}_{-340}$ & $0.0^{+1.59}_{-0.0}$ & HI, Mg II & 1,2 & (Aur) \\ 
 239960A & GJ 860A  & 104.7 & --0.0 & 4.01 & $-0.14\pm0.67$ & $4780^{+2080}_{-1780}$ & $2.24^{+0.42}_{-0.51}$ & D I, Mg II & 1 & (Eri) \\
 GJ 873 & EV Lac             & 100.6 & --13.1 & 5.0   &  $4.44\pm0.43$ & $3030^{+670}_{-610}$ & $2.21^{+0.22}_{-0.23}$ & D I, Mg II & 1 & (LIC)\\
 36395 & GJ 205  & 206.9 &--19.4 & 5.70 & $17.31\pm0.19$ & $4200^{+680}_{-680}$ & $2.37^{+0.24}_{-0.24}$ & H I, D I, Mg II & 1, 2 & unassigned \\  
 36395 & GJ 205  & 206.9 &--19.4 & 5.70 & $21.93\pm0.16$ & $4500^{+1200}_{-1200}$ & $3.64^{+0.27}_{-0.27}$ & H I, D I, Mg II & 1, 2 & (LIC) \\
  & GJ 588   & 332.7 & 12.1 & 5.92 & $-26.58\pm0.56$ & $6730^{+700}_{-720}$ & $3.37^{+0.32}_{-0.31}$ & H I, Mg II & 1,2 & G \\
  & YZ CMi   & 215.9 & 13.5 & 5.99 & $18.11\pm0.31$ & $8930\pm260$ & $0.0^{+1.18}_{-0.0}$ & H I, Mg II, Fe II & 1,2 & LIC \\
  & YZ CMi   & 215.9 & 13.5 & 5.99 & $21.73\pm0.53$ & $8790\pm230$ & $1.59^{+0.57}_{-0.77}$ & H I, Mg II, Fe II & 1,2 & (Aur) \\
191408 & GJ 783 &  5.2 & --30.9 & 6.01 & $-25.9\pm 0.42$ & $7700^{+1500}_{-1500}$ & $3.05^{+0.86}_{-0.86}$ & D I, Mg II & 1 & Mic\\
152751 & GJ 644B  & 11.0 & 21.1 & 6.20 & $-26.10\pm0.59$ & & & D I, Mg II & 2 & (Mic) \\
79210 & GJ 338A  & 164.9 & 42.7 & 6.33 & $12.58\pm0.09$ & $6650^{+970}_{-1020}$ & $2.29^{+0.61}_{-0.55}$ & D I, Mg II, Fe II & 1,2 & LIC \\
115617 & 61 Vir.    & 311.9 & 44.1  & 8.57 & $-16.7\pm0.47$ & $6600^{+4800}_{-4400}$ & $1.60^{+0.83}_{-1.60}$ & D I, C II, O I & 1 & (NGP)\\
23249 & $\delta$~Eri & 198.1 & --46.0 & 9.04 & $20.29\pm 0.43$ & $3650^{+1200}_{-1110}$ & $2.25^{+0.19}_{-0.22}$ & D I, C II, Mg II & 1 & LIC\\
37394   & GJ 211 & 158.4 & 11.9 & 12.3 & $17.2\pm0.27$ & $5600^{+2200}_{-2500}$ & $1.00^{+0.71}_{-1.00}$ & D I, C II, Mg II & 1 & LIC\\
166 & HR 8 & 111.3 &         --32.8 & 13.7 & $6.50\pm0.44$ & $4520^{+2110}_{-1880}$ & $4.39^{+0.33}_{-0.34}$ & D I, C II, Mg II & 1 & LIC\\
72905 & $\pi^1$~UMa & 150.6 & 35.7 & 14.5 & $12.91\pm 0.43$ & $2450^{+740}_{-660}$ & $2.47^{+0.10}_{-0.11}$ & D I, Mg II & 1 & LIC\\
142373 & $\chi$~Her & 67.7 & 50.3 & 15.8 & $-12.69\pm0.16$ & $4700^{+1200}_{-1100}$ & $1.36^{+0.31}_{-0.41}$ & D I, Mg II, Fe II & 1 & (Mic)\\
43162 & GJ 3389 & 230.9 & --18.5 & 16.7 & $17.1\pm0.43$ & $3200^{+2680}_{-2330}$ & $5.76^{+0.25}_{-0.26}$ & D I, Mg II & 1 & LIC\\
165185 & GJ 702.1 & 356.0 & -07.3 & 17.2 & $-26.9\pm 0.43$  & $8610^{+4150}_{-3340}$ & $0.0^{+1.76}_{-0.00}$ & D I, Mg II & 1 & (G)\\
116956 & SAO 28753 & 113.7 & 59.5 & 21.7 & --0.25 & $5160^{+3420}_{-2680}$ & $1.79^{+0.69}_{-1.37}$ & D I, Mg II & 1 & (LIC)\\
116956 & SAO 28753 & 113.7 & 59.5 & 21.7 & 7.7  & $2540^{+2450}_{-1830}$ & $1.66^{+1.21}_{-1.66}$ & D I, Mg II & 1 & unassigned\\ 
59967   &  GJ 3446 & 250.5 & --09.0  & 21.8 & $9.32\pm 0.02$ & $12400^{+2900}_{-2700}$ & $0.00^{+1.9}_{-0.0}$ & D I, C II, O I & 1 & Blue\\
 97334   & GJ 417  & 184.3 & 67.3 & 22.6 & $3.34\pm0.47$ & $11000^{+2400}_{-3000}$ & $0.0^{+1.4}_{-0.0}$ & D I, C II, O I & 1 & LIC\\
 73350.  & GJ 9273  &  232.1 & 20.0 & 24.3 & 11                   & $7900^{+3200}_{-3200}$ & $0.68^{+1.67}_{-0.68}$ & D I, Mg II & 1 & LIC \\
 82210 & DK UMa & 142.5 &  38.9 & 32.4 & $9.41\pm0.61$ & $4220^{+460}_{-430}$ & $1.86^{+0.15}_{-0.18}$ &  D I, Mg II & 1 & (LIC)\\
129333 & EK Dra & 105.5 & 49.0 & 34.6 & $-2.43\pm 0.32$ & $2580^{+280}_{-250}$ & $3.13^{+0.05}_{-0.05}$ & D I, C II, Mg II &1\ & (LIC)\\
93497 & $\mu$~Vel & 283.0 & 08.6 & 35.9 & $-4.38\pm 0.43$ & $3470^{+580}_{-590}$ & $3.12^{+0.11}_{-0.10}$ & D I, Mg II & 1 & G\\
128787 & SAO 182739 & 331.0 & 30.2 & 42.3 & $-20.90\pm0.62$ & $6580^{+4500}_{-3680}$ & $3.00^{+0.11}_{-0.10}$ & D I, Mg II& 1 & Gem\\
209458 & V376 Peg & 76.8 & --28.5 & 48.1 & $-29.17\pm0.94$ & $8500^{+3280}_{-2850}$ & $0.00^{+3.02}_{-0.00}$ & D I, Mg II & 1 & unassigned\\
\hline\hline
\end{tabular}
\end{center}
References: (1) This paper;  (2) \cite{Wood2021}.\\
\end{table}

\section{Analyzing the distributions of cloud properties}

Until recently, interstellar clouds in the LISM have been characterized by the mean properties measured for all sight lines passing though the cloud consistent with the cloud's velocity vector. With the availability of 100 velocity components passing through the LISM and 37 of these passing though the LIC, we can begin to do statistical analyses of the data. In the following analyses, we include only those velocity components for which the temperatures are at least two times larger than the measurement errors, because the software may not fully separate thermal from turbulent velocity and there could be unresolved velocity components that would appear as extra thermal broadening. For the entire data set 84 of the 100 velocity components meet this $2\sigma$ criterion, and for velocity components passing through the LIC 36 of the 37 meet this criterion. In this analysis we include sight lines out to stars only within 100~pc. 
We search for trends in physical properties by asking specific questions.

\subsection{Are the temperatures and turbulent velocities within a cloud roughly constant or variable?}

An important question is whether the cloud temperatures and turbulent velocities within a cloud are roughly constant or distributed in a random or systematic manner.  If these properties are random, then their distributions could be Gaussian distributed with perhaps a few outliers. With the 84 sight lines now evaluated for temperature and turbulent velocity, we can address this question. Figure~\ref{binvsTandTLIC} (left) shows the number of temperature measurements in each 1000~K temperature bin. This bin size is appropriate given the typical measurement uncertainties of 500--2,000~K. The weighted mean gas temperature is $6,838\pm2,455$~K, and the solid line in the plot is a Gaussian fit to the data. The fit is reasonably good, although there are  6 outliers in the temperature range 12,000--13,000~K. The errors are weighted dispersions about the mean \citep[cf.][]{Redfield2004b}. Table~1 lists the weighted mean temperatures and dispersions measured in different ways.  Figure~\ref{binvsTandTLIC} (right) shows the distribution of temperatures for the 34 sight lines that pass through the LIC. While there are fewer sight lines in this plot, the distribution, mean value, and dispersion of the temperatures are similar to the plot that included all sight lines. These data show that the LIC and the other warm clouds in the CLIC have similar wide distributions of internal temperatures.

Figure~\ref{binvsTurbandTurbLIC} (left) shows the distribution of turbulent velocities plotted with a bin size of 0.5~km~s$^{-1}$ as typical measurement uncertainties are in the range 
0.2--1.5~km~s$^{-1}$. The weighted mean turbulence is $2.54\pm 1.18$~km~s$^{-1}$, and the solid line Gaussian in the figure is also a good fit to the data. The distribution of turbulent velocities for the sight lines that pass through the LIC (Figure~\ref{binvsTurbandTurbLIC} (right)) has similar properties. We conclude that wide ranges of temperatures and turbulent velocities characterize the gas in the nearby warm clouds. As previously noted by \cite{Redfield2004b}, the turbulent velocities are subsonic and the highest turbulent velocities may result from unresolved velocity components. 

\setcounter{figure}{1}
\begin{figure*}[htb!]
\includegraphics[width=9.0cm]{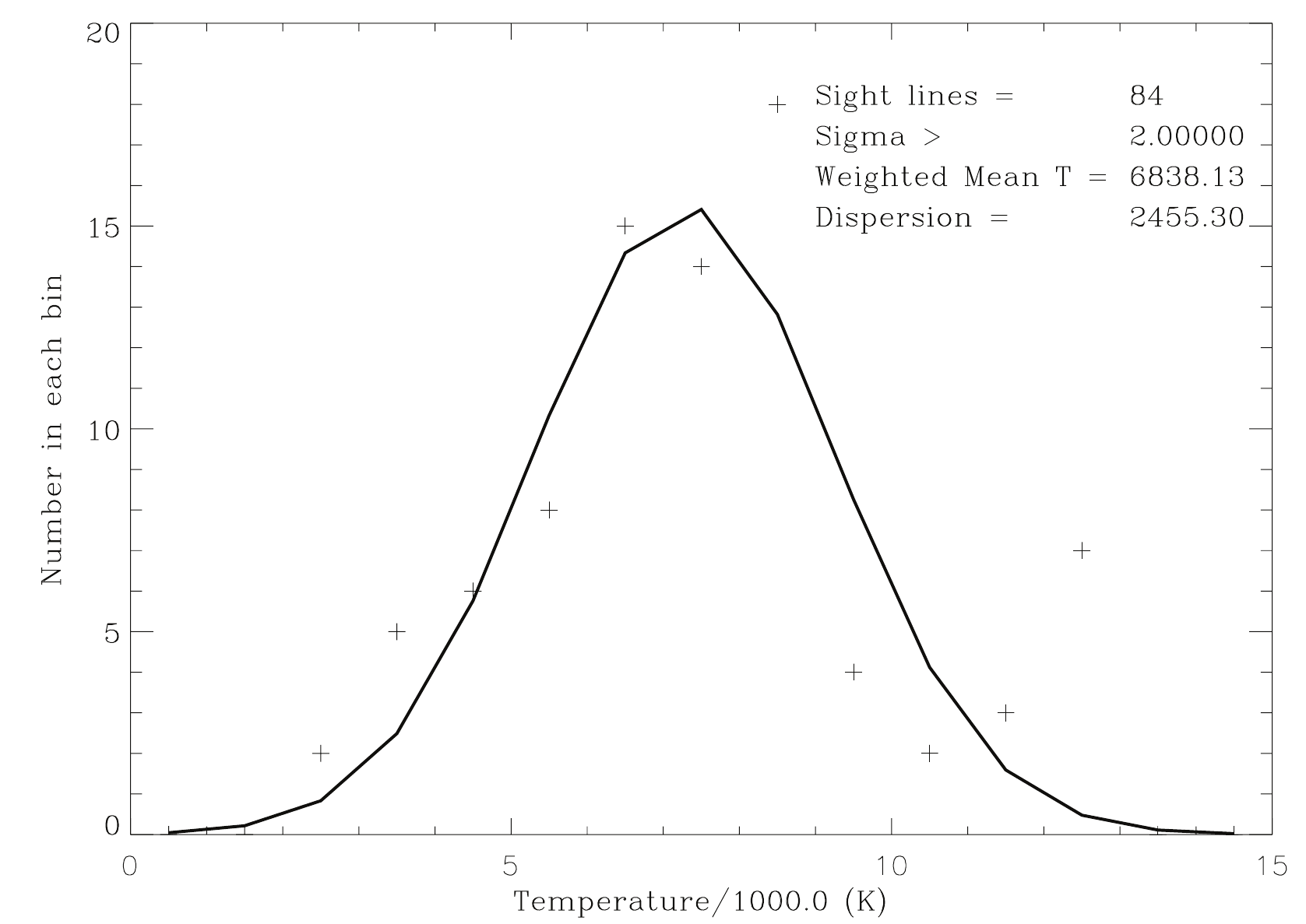}
\includegraphics[width=9.0cm]{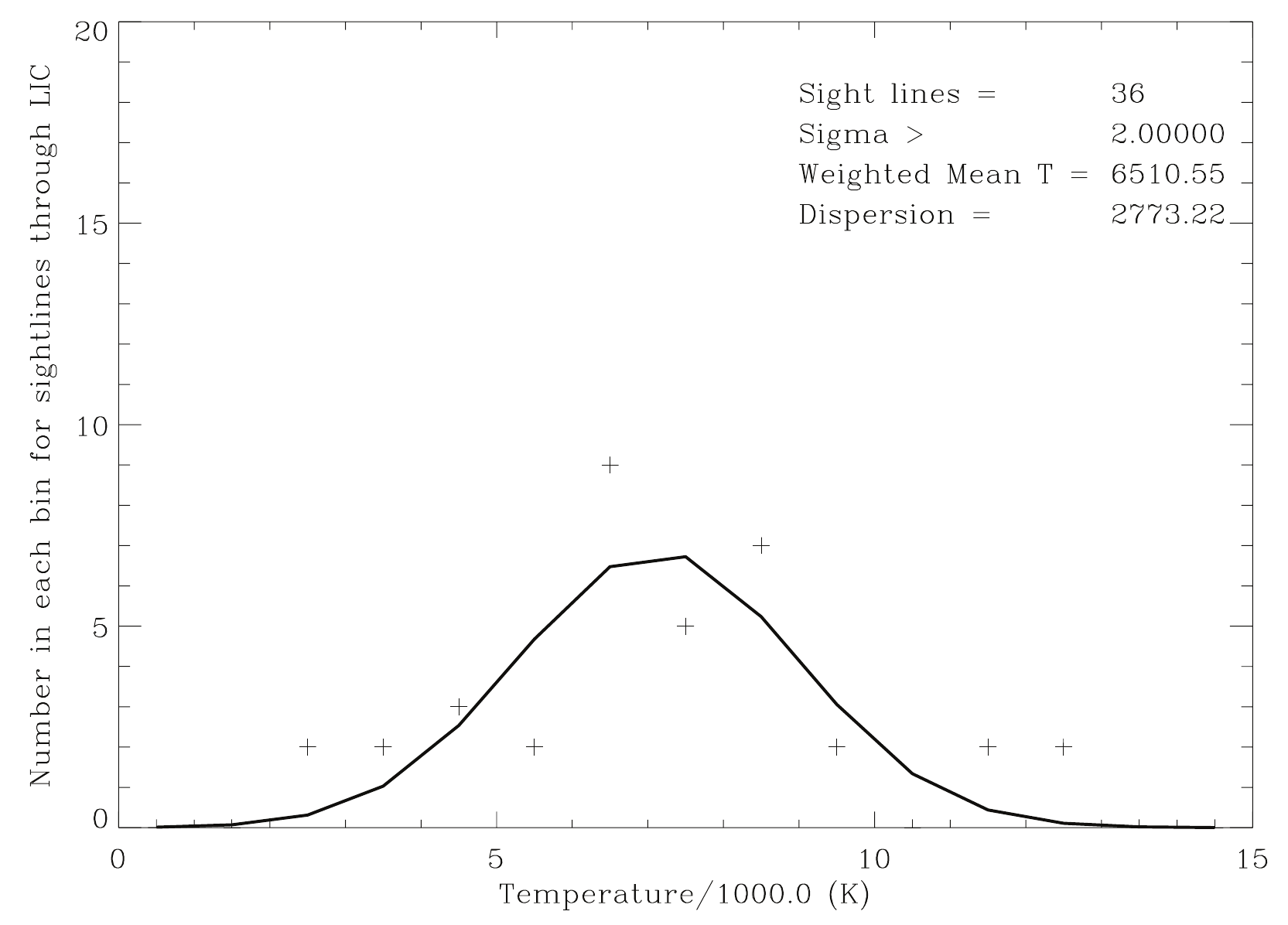}
\caption{Distributions of sight line temperatures for the full data set ({\bf left}) and for only the sight lines that pass through the LIC ({\bf right}). Temperatures are binned in 1000~K intervals. The solid curves are Gaussian fits to the weighted data points. \label{binvsTandTLIC}}
\end{figure*}

\begin{figure*}[htb!]
\includegraphics[width=9.0cm]{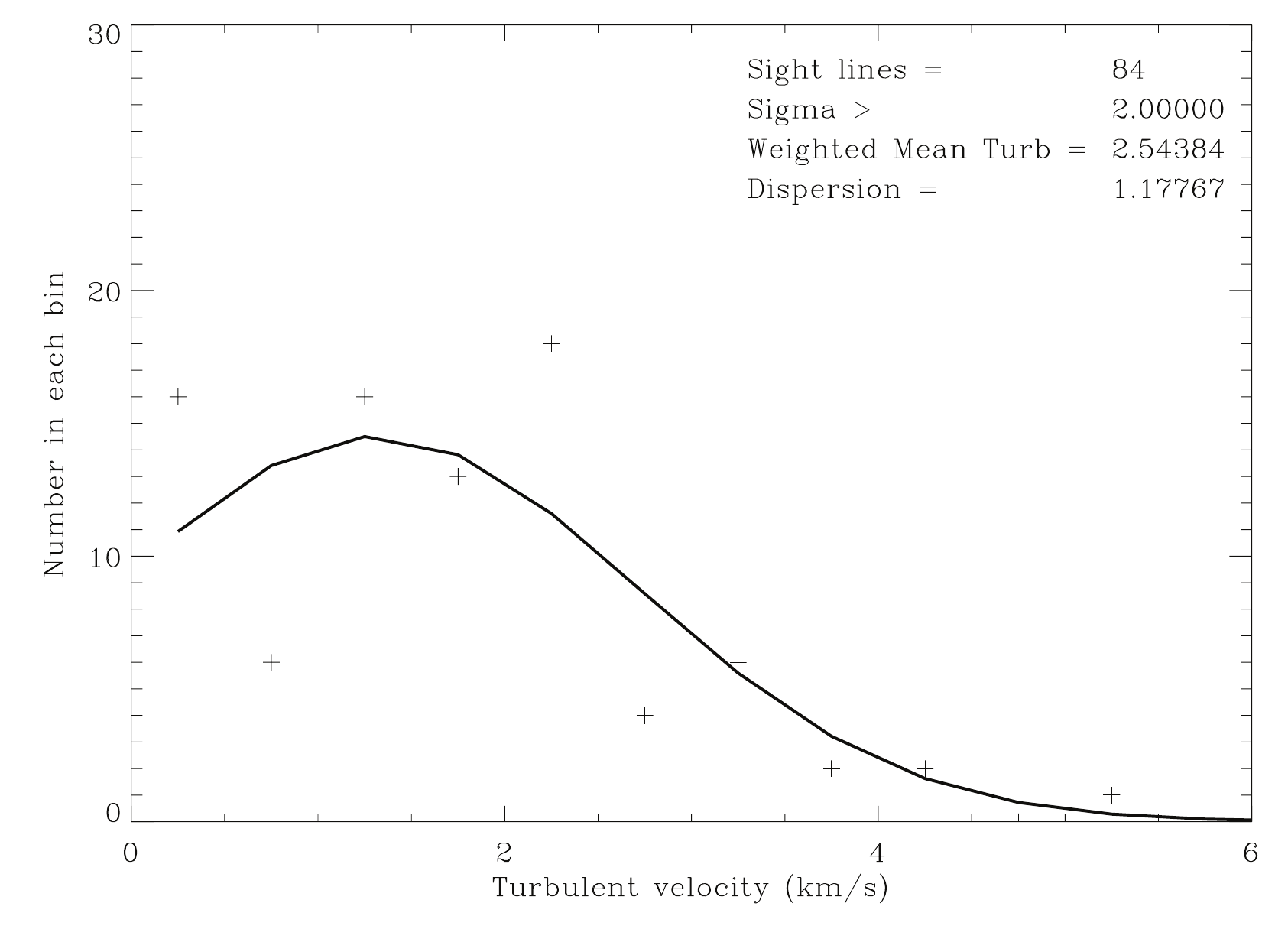}
\includegraphics[width=9.0cm]{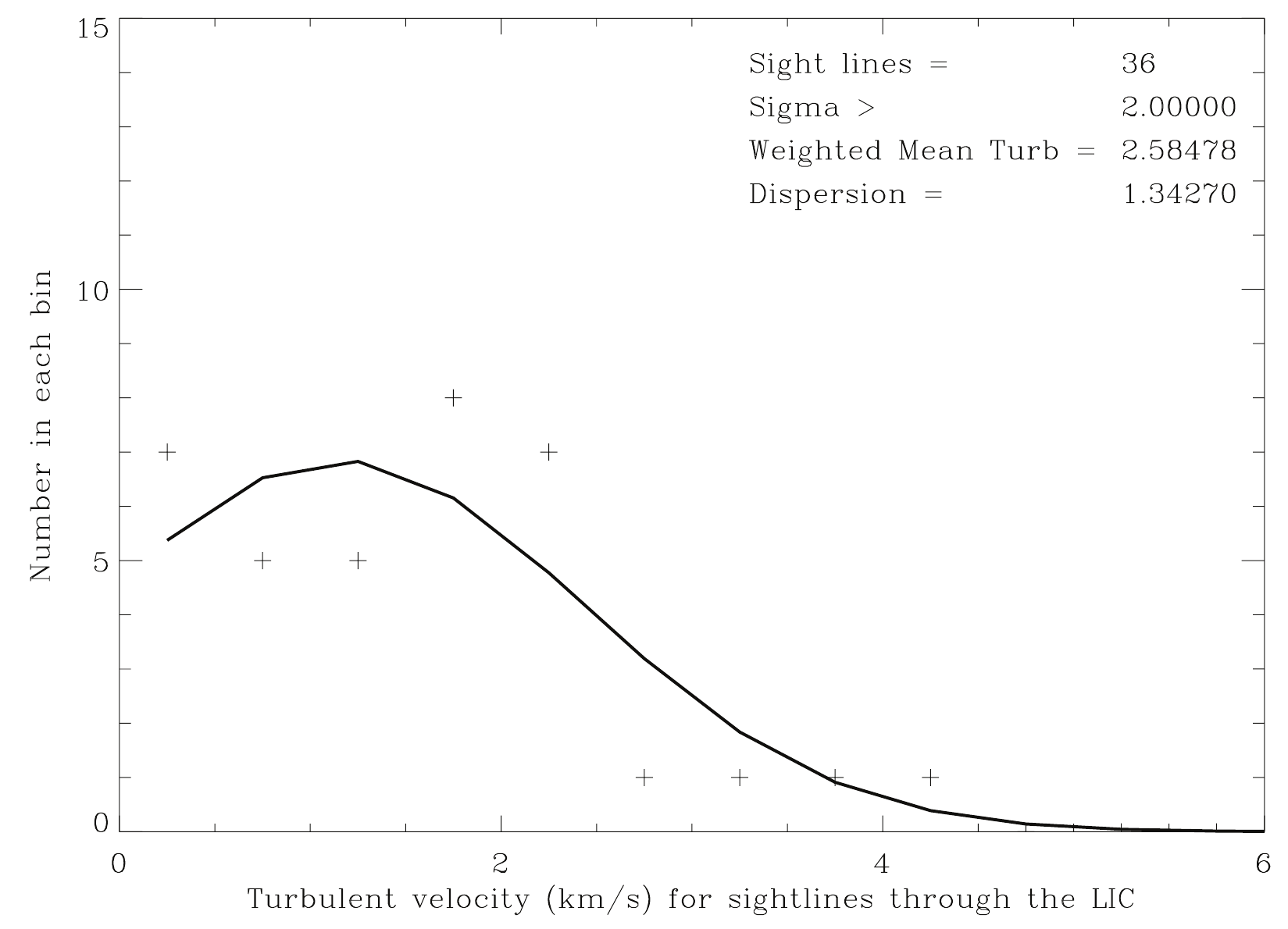}
\caption{Distributions of sight line turbulent velocities for the full data set ({\bf left}) and for only the sight lines that pass through the LIC ({\bf right}). Turbulent velocities are binned in 0.5 km/s intervals. The solid curves are Gaussian fits to the weighted data points. \label{binvsTurbandTurbLIC}}
\end{figure*}

\begin{figure*}[htb!]
\includegraphics[width=9.0cm]{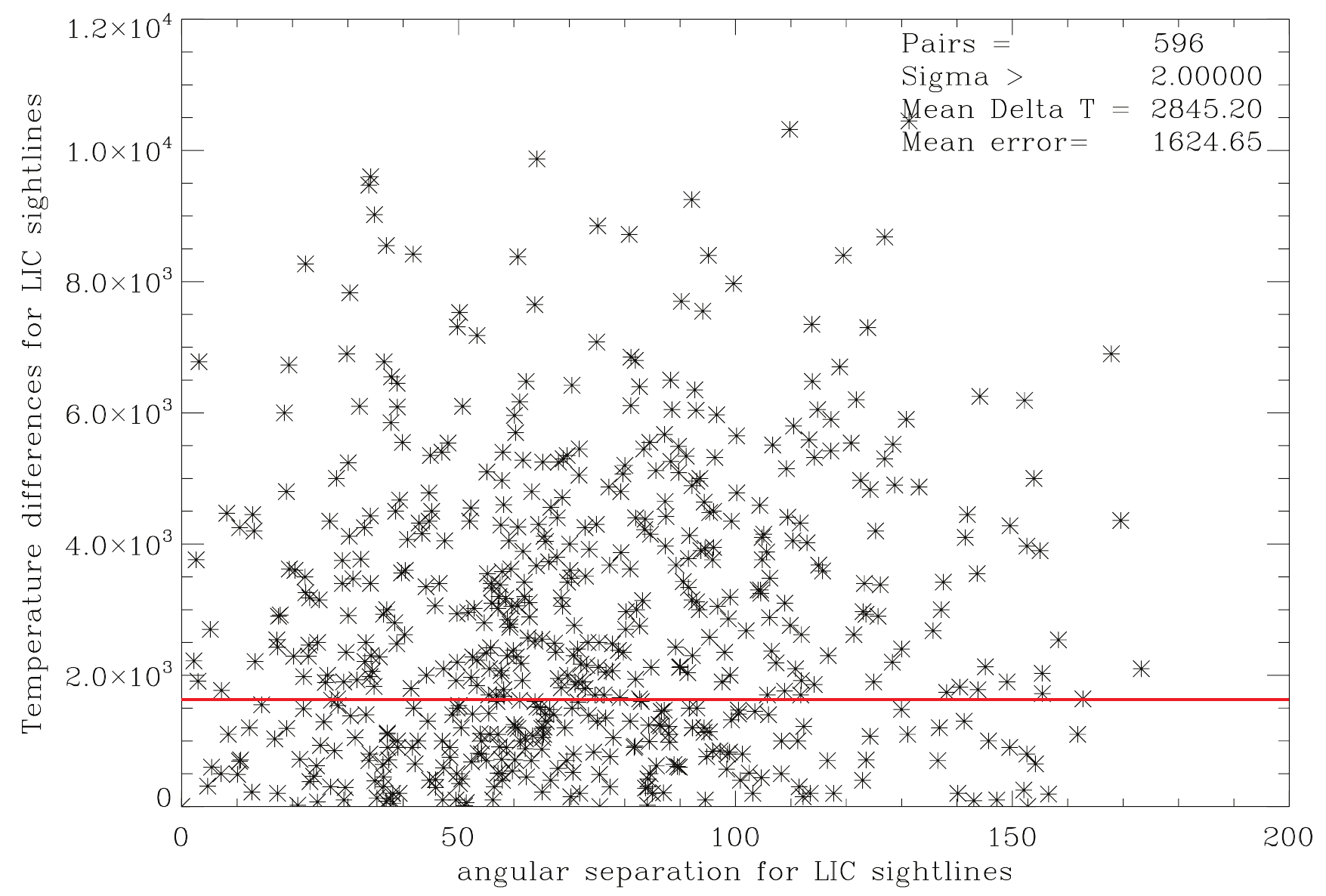}
\includegraphics[width=9.0cm]{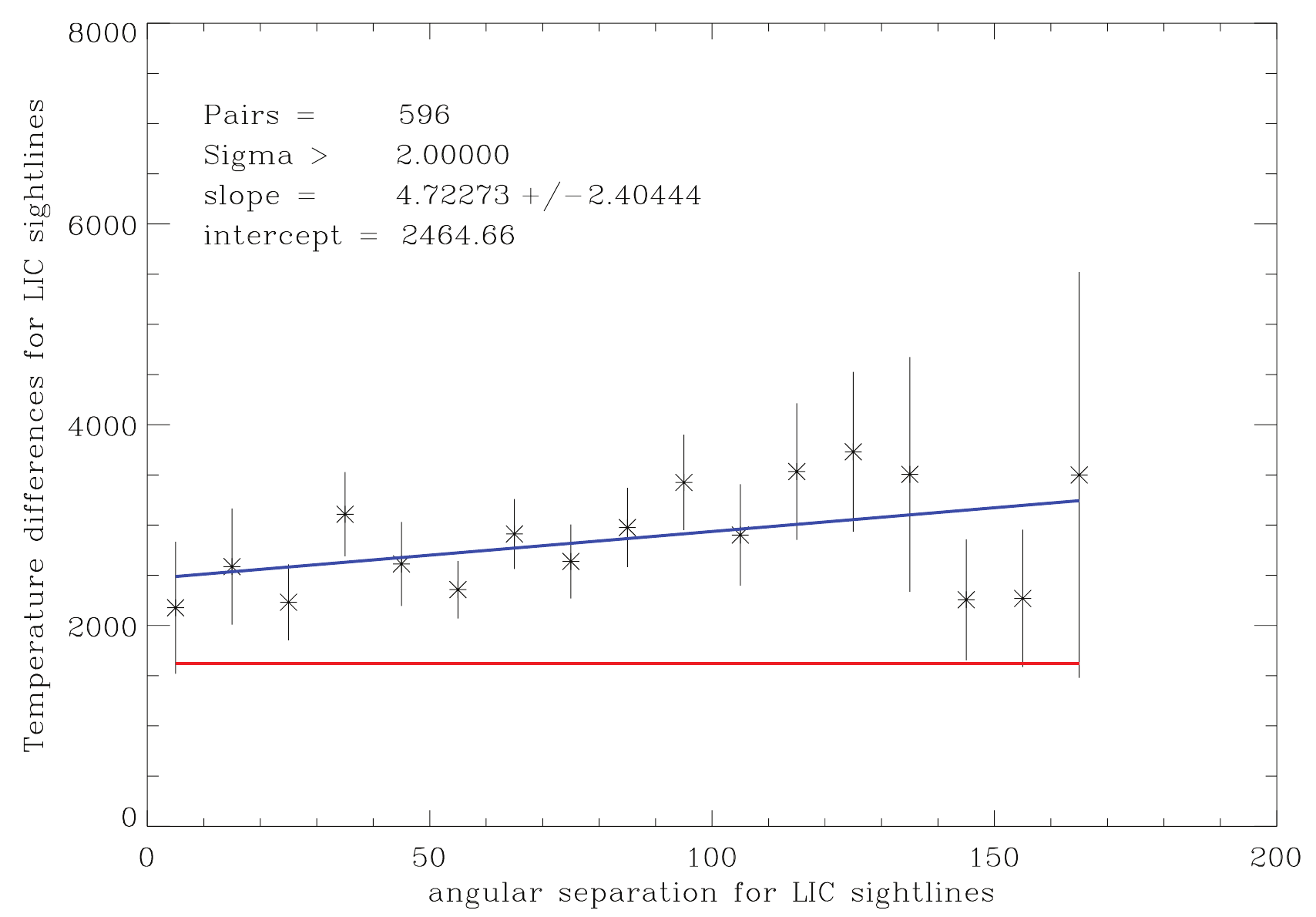}
\caption{{\bf Left:} Temperature differences for sight line pairs through the LIC as a function of angular separation between the sight lines. The horizontal red line is the mean uncertainty of temperature measurements. {\bf Right:} Mean temperature differences for sightline pairs through the LIC in $10^{\circ}$ angular bins. The vertical lines show the variance of the temperature differences in each angular bin and the star symbols are the mean value in each bin. The blue line is a least squares fit to the mean temperature differences as a function of angular separation.}
\label{TdiffLICsepvsangle}
\end{figure*}

Another method for testing whether the temperatures and turbulent velocities within the LIC are constant or have variable properties is to compare the properties of pairs of sight lines as a function of their angular separation. For the 36 sight lines traversing the LIC there are 36x35=1260 pairs but only 596 unique pairs after subtraction of pairs for two clouds in the sight line to the same star and pairs that are sampled twice by the search software. Figure~\ref{TdiffLICsepvsangle} (left) shows the temperature differences for these pairs as a function of angular separation between the sight lines. The mean value of these temperature differences is 2845~K, but there are many temperature differences exceeding 6000~K. Since the mean value of the individual measurement uncertainties for these sight lines is 1625~K, most of these sightline pairs have temperature differences that well exceed the measurement errors. 

Figure~\ref{TdiffLICsepvsangle} (right) plots the same data in angular separation bins $10^{\circ}$ wide. The blue line is a least-squares fit to the mean values in each bin. The line slope is 1.96 times its error, indicating a significant increase in the mean  temperature differences between zero angular separation separation  (2465~K) and those at the widest separation. This result indicates significant temperature differences on all angular separation scales, even at the smallest separations.

\subsection{What is the length scale for the temperature inhomogeneities in the LIC?} 

There are many studies of inhomogeneous properties in the interstellar medium. See, for example, the comprehensive review by \cite{Stanimirovic2018}. Very small scale structures of neutral cold gas measured from absorption lines of neutral hydrogen and molecules toward pulsars and other sources reveal tiny scale atomic structures (TSASs) as small as tens of au. Very small structures of ionized gas in the warm diffuse interstellar medium observed from pulsar scintillations also show tiny structures with sub-au scales. These small spatial scales for both neutral and ionized gas refer to density and magnetic field fluctuations and are based on measurements through long sight lines through the Galaxy. With the available data, we can now measure the length scales for temperature fluctuations in the immediate environment of the Sun, including the LIC and other clouds, and what these fluctuations suggest could be the timescale of externally driven changes in the heliosphere. 

The least squares fit in Fig~\ref{TdiffLICsepvsangle} predicts that the mean temperature difference for sight lines with the smallest angular separations should be about 2,465~K, but the mean temperature measurement error is only 1625~K. For stars in our data set, the smallest angular scale of $2^{\circ}$.2 is for the Procyon-YZ CMi sight line pair, which has a temperature difference of 2220~K, but the measurement uncertainty for this pair is only 695~K, a factor of 3.2 smaller.  Table~4 lists the four sight line pairs with the closest angular separations. We estimate the path length $d$(LIC) through the LIC from the neutral hydrogen column densities and $n$(H~I)=0.20~cm$^{-3}$. The separation $s$ of the sight lines half way through the LIC is then $s=206265\times(d($LIC$)/2)\tan(\theta)$~au, where $\theta$ is the angular separation of the sight lines. The separations range from 5100~au to 17,360~au for these close pairs. These separations could be upper limits to the true inhomogeneity length scale, because there could be unmeasurable but significant temperature variations within the sight lines to even the closest stars and between the stars with the smallest angular separations. Since the Sun moves through the LIC at 5.1 au/yr, the heliosphere could see changes in local interstellar properties within 1000 years. We conclude that the linear scale for significant temperature differences in the LIC is at least as small as 5,100~au. 

This technique for estimating length scales is similar to that described by \cite{Spangler2001}, who compared radiation measures as a function of angle between different sight lines. Radiation measures are proportional to the integral of electron densities along a line of sight. There is a break in the difference between radiation measures for sight lines with angular separations less than or about $0^{\circ}.1$, corresponding to a length scale for these longer path lengths of 3.6~pc.

\begin{table}
\caption{Sight line separations halfway through the LIC  for the closest pairs} 
\begin{center}
\begin{tabular}{lcccccccc}
\hline\hline
Angle & Star1 & Star2 & $\Delta T$ & Uncertainty & $\Delta T$/Uncertainty & log N(H~I) & $d$(LIC) & Separation\\ \hline
$2^{\circ}.20$ & Procyon & YZ CMi & 2220 K & 695 K & 3.2\ & 17.9, 17.89 & 1.29 pc& 5,100 au\\
$2^{\circ}.62$ & $\epsilon$~Eri & $\delta$~Eri & 3760 K & 1435 K &2.6 & 17.88, 17.88 & 1.29 pc& 6,075 au\\
$2^{\circ}.94$ & Procyon & GJ 273 & 1910 K & 733 K & 2.6 & 17.9, 17.86 & 1.23 pc& 6,510 au\\
$3^{\circ}.12$ & HD 166 & PW And & 6780 K & 2724 K & 2.5 & 18.46, 18.1 & 3.09 pc& 17,360 au\\ 
\hline\hline
\end{tabular}
\end{center}
\end{table}

\subsection{Are temperatures and turbulent velocities correlated?}

\begin{figure}[htb!]
\includegraphics[width=9.0cm]{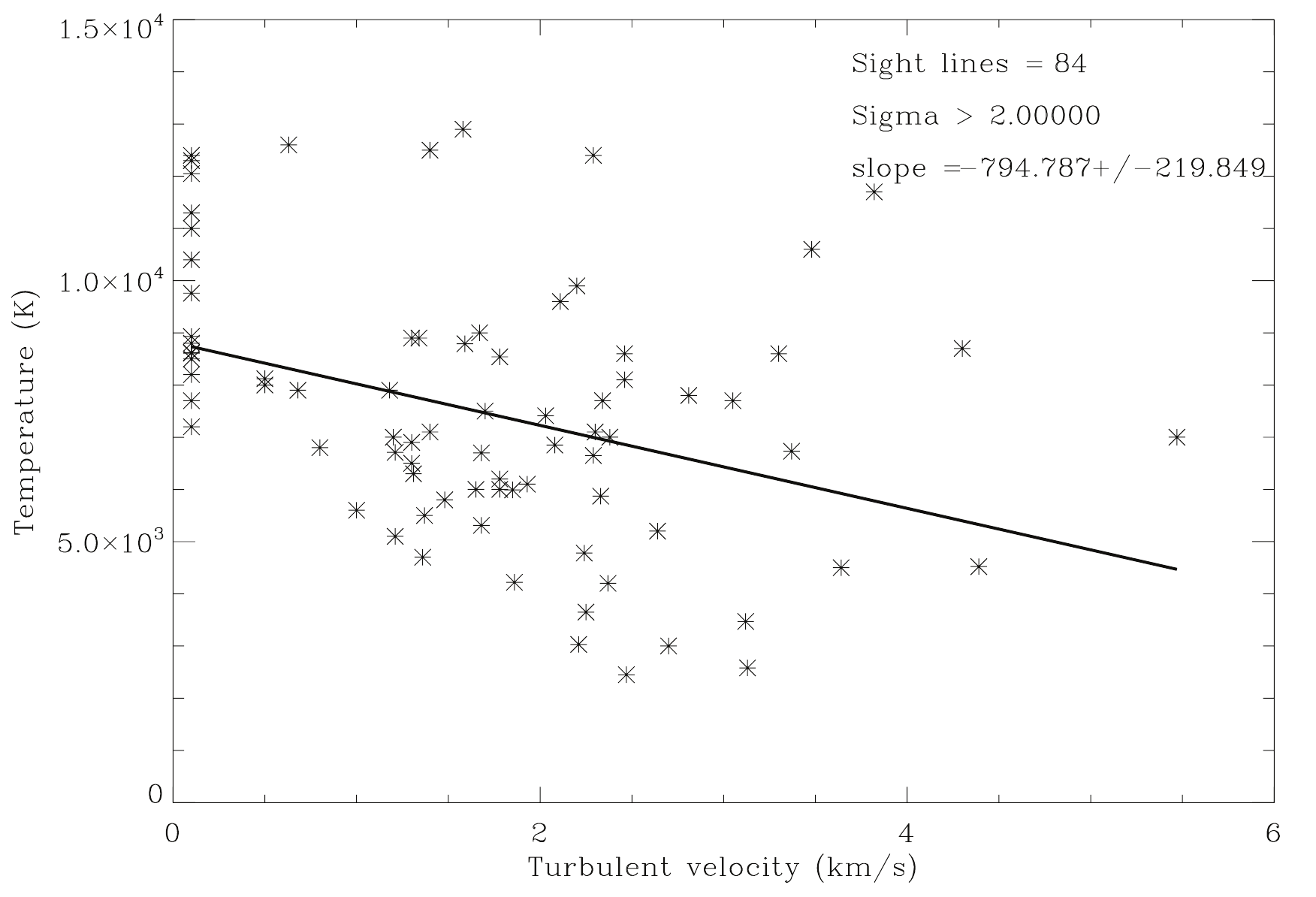}
\includegraphics[width=9.0cm]{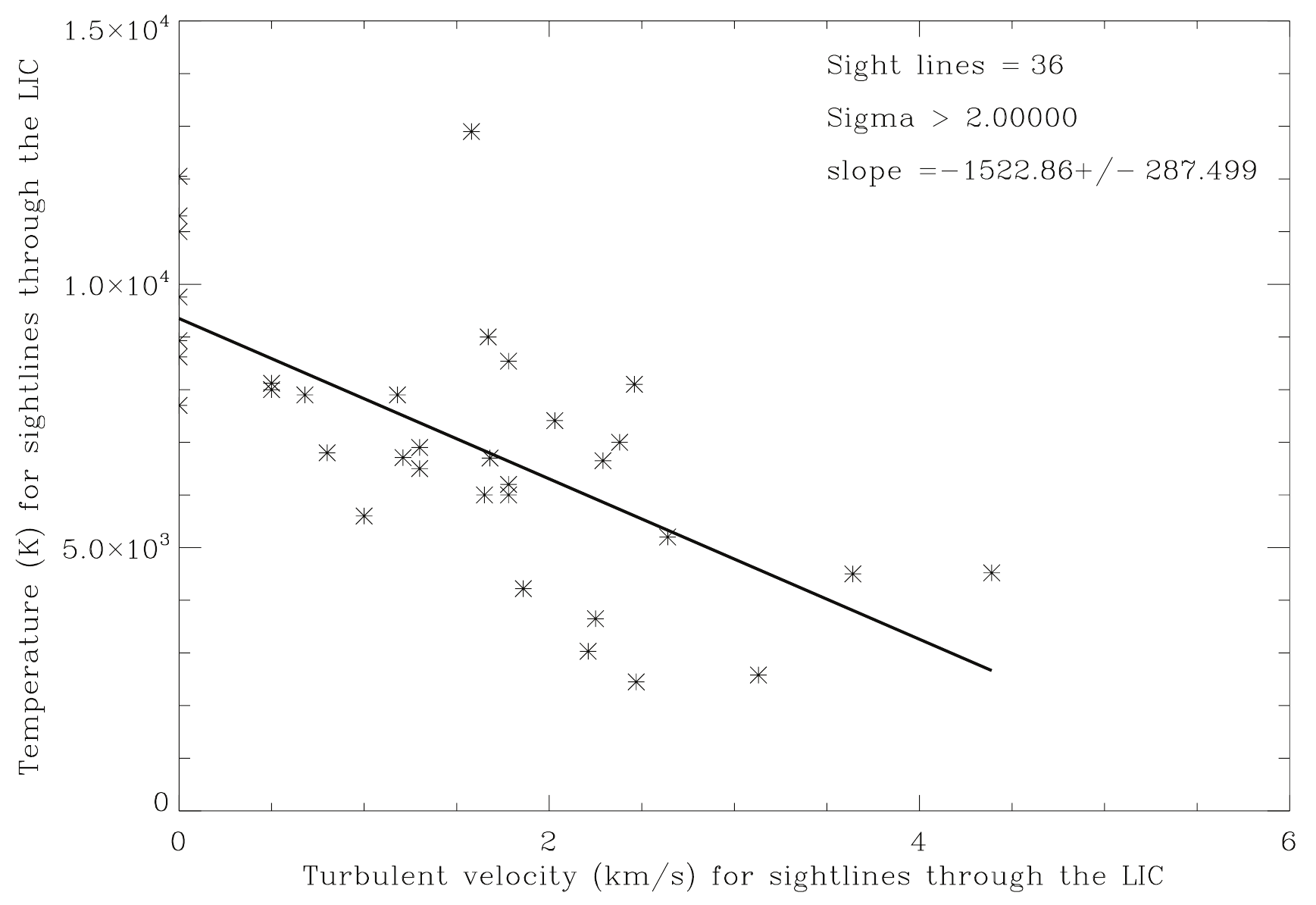}
\caption{Plots of sightline temperatures vs turbulent velocities for the full data set ({\bf left}) and the LIC data set ({\bf right}). The solid lines are least-squares linear fits to the data. \label{TvsTurb}}
\end{figure}

Figure~\ref{TvsTurb} (left) plots temperature vs turbulent velocity for all 84 velocity components. Although there is a large scatter in the data, there is a clear trend of decreasing temperature with increasing turbulent velocity. For the full data set, the  least-squares linear fit in the form $T=A+Bv_{\rm turb}$, where $A=8816\pm449$~K and $B=-795\pm220$, with the slope 3.6 times its error. For the LIC data (Figure~\ref{TvsTurb} (right)), the fit is $A=9350\pm 523$ with a steeper slope $B=-1523\pm 287$ that is 5.3 times its error. This significant dependence of temperature on turbulent velocity could result in part from the absorption line fitting procedure, because the fit to low mass atoms (e.g., D~I) and high mass ions (e.g., Mg~II and Fe~II) depends on the sum of thermal and turbulent broadening. For example, a positive (or negative) error in the temperature measurement can be partially compensated by negative (or positive) error in the turbulent velocity measurement. The presence of unresolved velocity components would significantly increase the measured turbulent velocities with only a small effect on the temperature measurements. Since the effect of unresolved velocity components is to produce the largest turbulent velocities, the correction of these data points would produce an even steeper correlation. The correlation could well be real and not just an artifact of the measurement technique or inadequate spectral resolution. A possible explanation would be the conversion of turbulent energy to heat. Table~5 summarizes the $A$ and $B$ coefficients for all of the linear fits. The underlined values of $B$ indicate significant trends.

\begin{table}
\caption{Linear Least-squares fits for $y=A+Bx$}
\begin{center}
\begin{tabular}{lccc}
\hline\hline
Plot for parameters y vs x & Data set ($2\sigma$) & $A$ & $B$\\ \hline
Temperature vs. turbulent velocity & All & $8816\pm449$ & \underline{$-795\pm220$}\\
Temperature vs. turbulent velocity & LIC & $9350\pm523$ & \underline{$-1523\pm287$}\\
Temperature vs. distance & All & $6969\pm379$ & \underline{$28.7\pm15.0$}\\
Temperature vs. angle from Galactic Center & All & $6804\pm614$ & $7.20\pm5.86$\\
Temperature vs. angle from $\epsilon$~CMa & All & $7704\pm627$ & $2.53\pm6.42$\\
Temperature vs. angle from $\epsilon$~CMa & LIC & $7342\pm1010$ & $-2.83\pm11.90$\\
Temperature vs. angle from inflow direction & All & $6669\pm635$ & $8.42\pm5.96$\\
Temperature vs. angle from LIC core & LIC & $6127\pm1032$ & $17.01\pm16.16$\\
Temperature vs, N(H~I) & All & $7657\pm13068$ & $-9.81\pm731.1$\\ 
Temperature vs, N(H~I) & LIC & $171434\pm19113$ & $-560\pm1067$\\ 
Temperature difference vs. angle from LIC core & LIC & $2465\pm236$ & \underline{$4.72\pm2.40$}\\ 
Turbulent velocity vs. distance & All & $1.764\pm0.180$ & $-0.00468\pm0.00713$\\
Turbulent velocity vs. angle from Galactic Center & All & $1.612\pm0.289$ & $0.000722\pm0.00276$\\
Turbulent velocity vs. angle from $\epsilon$~CMa & All & $1.952\pm0.291$ & $-0.00310\pm0.00298$\\
Turbulent velocity vs. angle from $\epsilon$~CMa & LIC & $1.239\pm0.444$ & $0.00289\pm0.00524$\\
Turbulent velocity vs. angle from LIC core & LIC & $1.596\pm0.463$ & $ -0.00228\pm0.00724$\\ 
Turbulent velocity vs. angle from inflow direction & All & $1.623\pm0.300$ & $0.000592\pm0.00281$\\
Turbulent velocity vs. N(H~I) & All & $-4.33\pm6.20$ & $0.335\pm0.347$\\
Turbulent velocity vs. N(H~I) & LIC & $-4.69\pm8.41$ & $0.339\pm0.470$\\
\hline\hline
\end{tabular}
\end{center}
\end{table}

\subsection{Do temperatures and turbulent velocities depend on the stellar distance?}

\begin{figure*}[htb!]
\includegraphics[width=9.2cm]{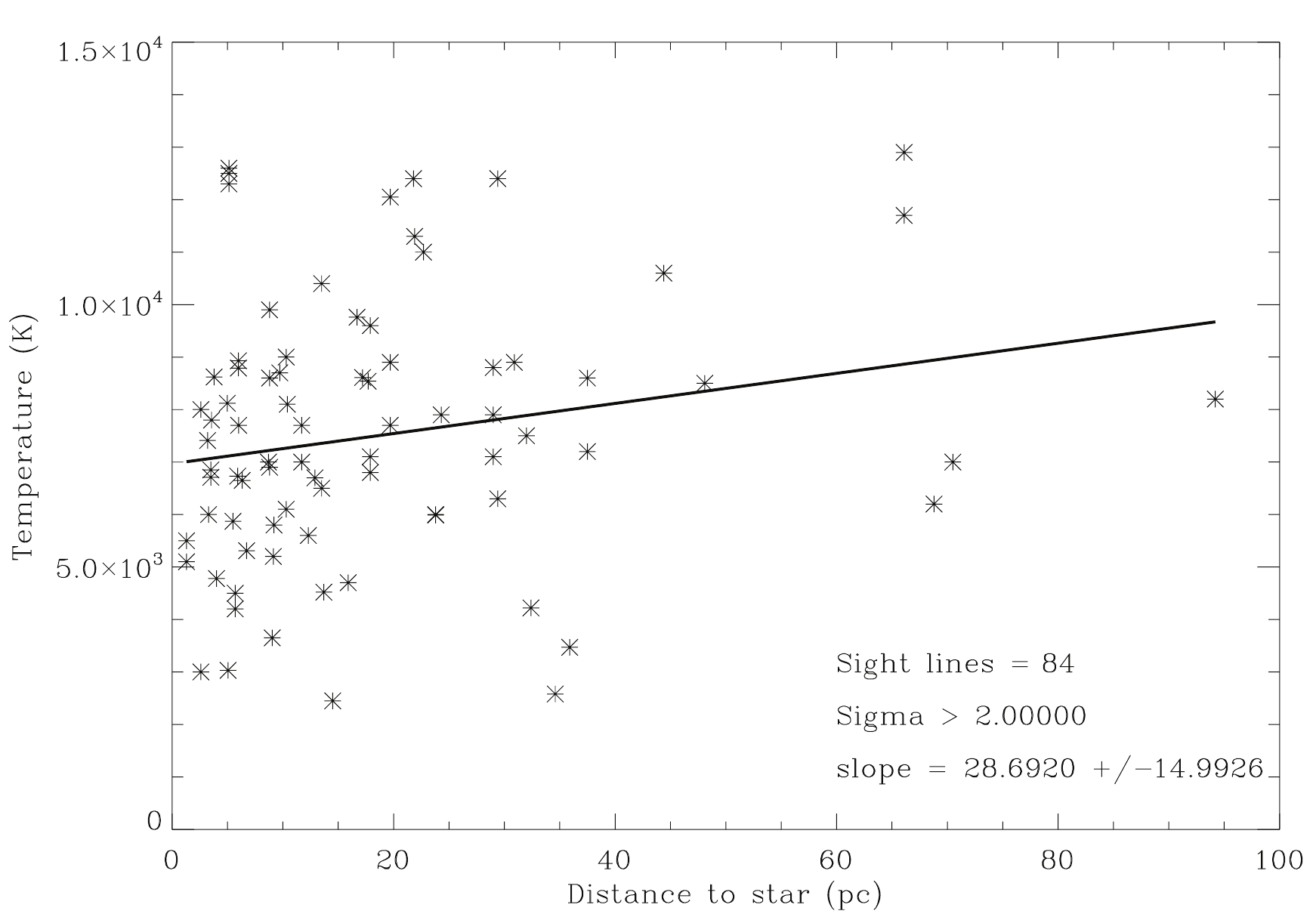}
\includegraphics[width=8.8cm]{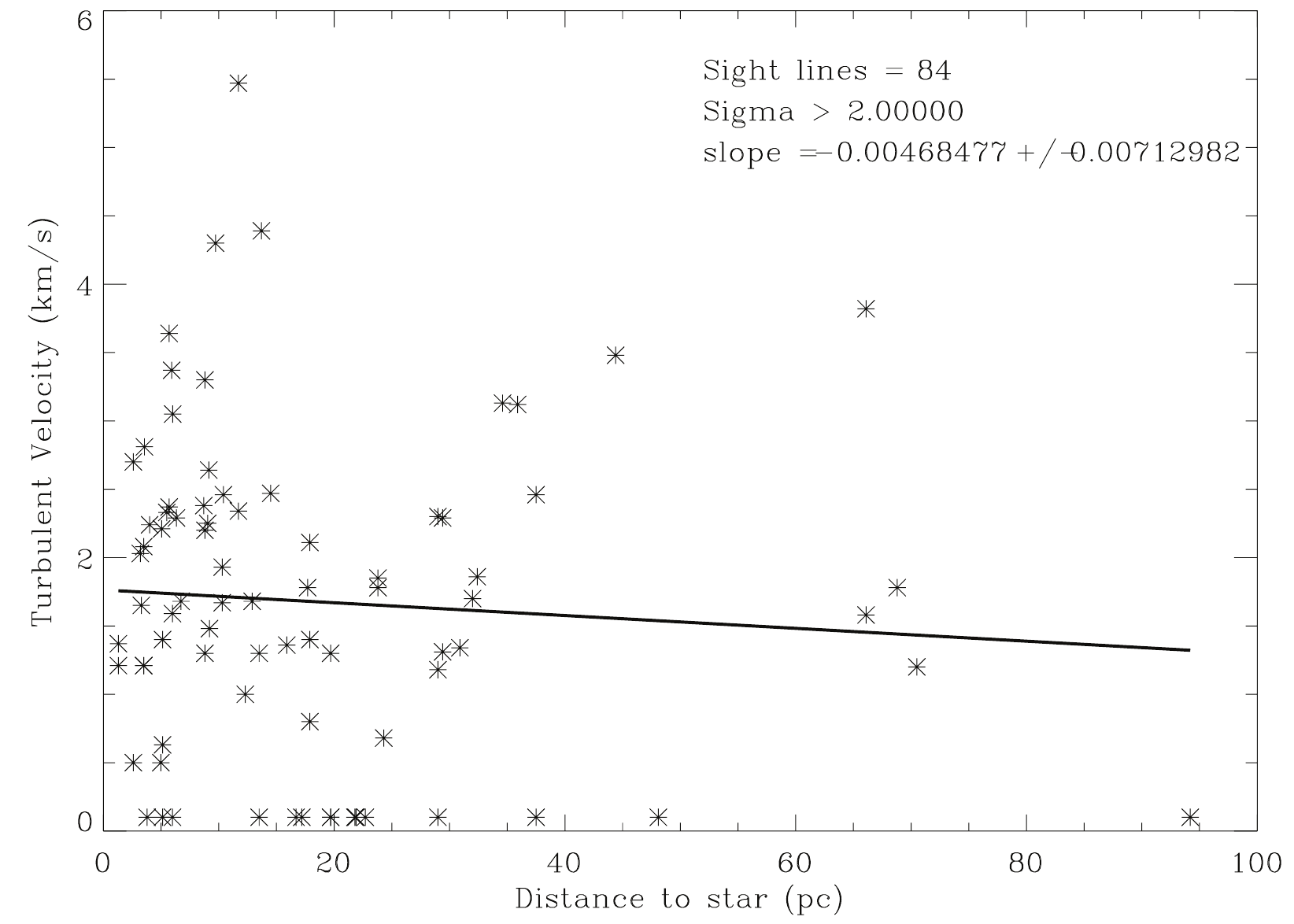}
\caption{Plots of sight line temperatures ({\bf left}) and turbulent velocities ({\bf right}) vs distance to the star. The solid lines are least-squares linear fits to the data. \label{TandTurbvsd}}
\end{figure*}

We searched for evidence of whether the distance to the star at the end of the sight line influences the properties of the intervening interstellar gas. Figure~\ref{TandTurbvsd} (left) plots temperature vs the distance to the star at the end of a given sight line. The absence of a significant trend in the data supports the assumption that  background stars only serve as illumination sources. The fit parameter for the LIC data in Figure~\ref{TandTurbvsd} (right) also shows no significant trend in the available data. A possible trend with distance to stars with very strong EUV emission is tested in Section 3.6. We can therefore combine sight lines of near and more distant stars when studying individual clouds. 

\subsection{Do temperatures and turbulent velocities depend on angle with respect to the Galactic Center and the inflow directions?}

\begin{figure*}[htb!]
\includegraphics[width=9.2cm]{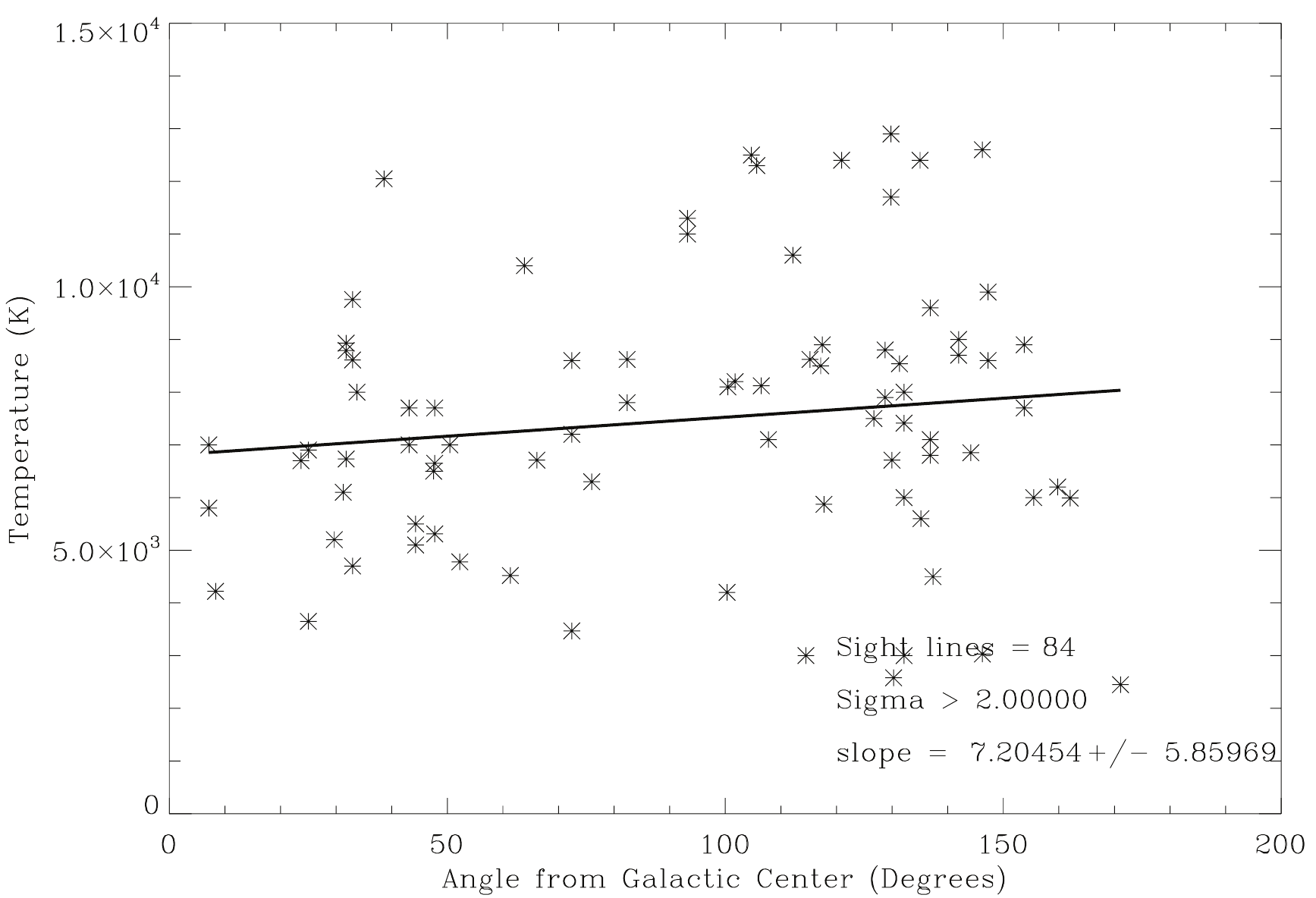}
\includegraphics[width=8.8cm]{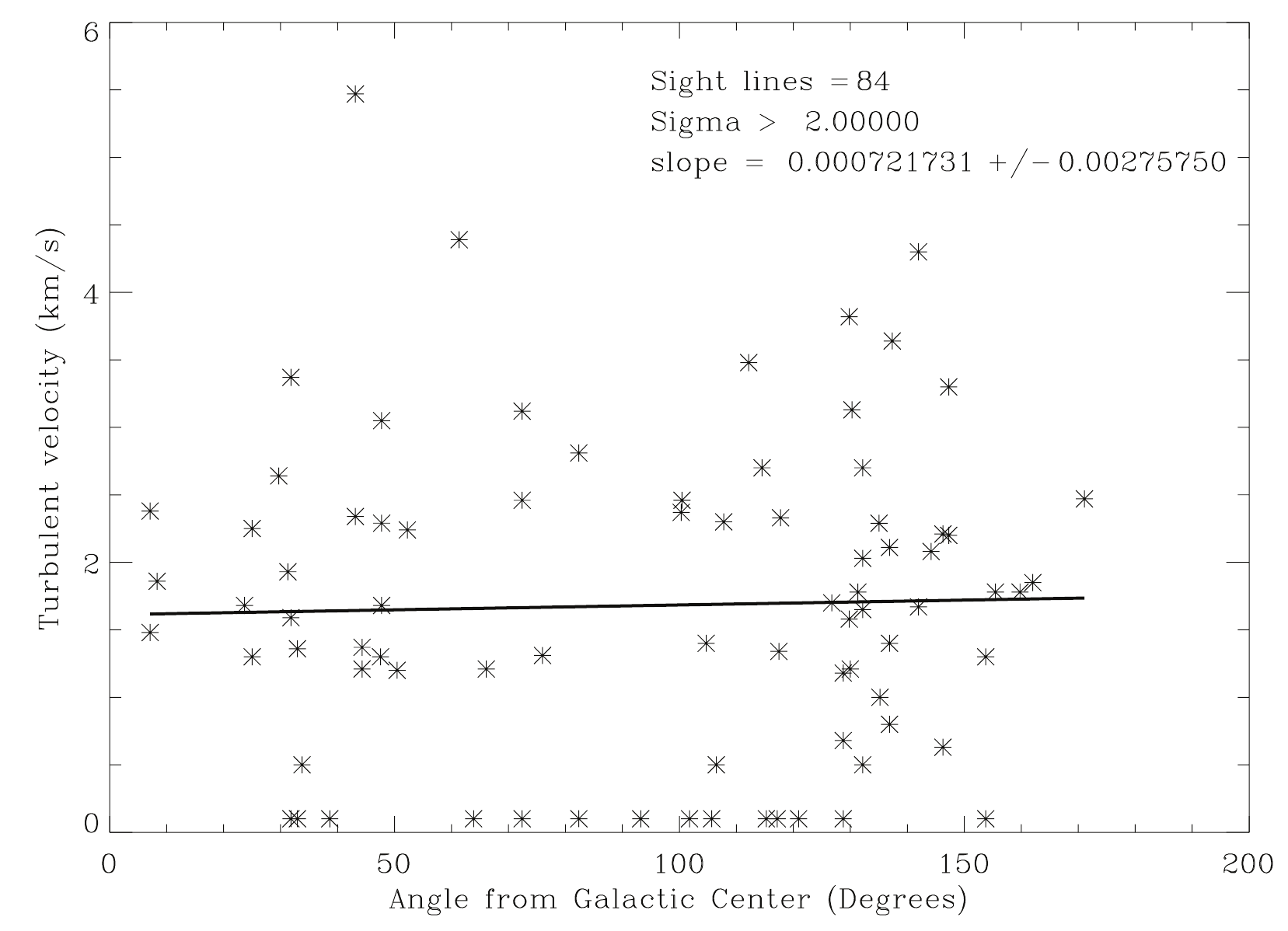}
\caption{Plots of sight line temperatures ({\bf left}) and turbulent velocities ({\bf right}) vs angle from the Galactic Center. The solid lines are least-squares linear fits to the data. \label{TandTurbvsGC}}
\end{figure*}

We next considered whether the temperature and turbulent velocity of local interstellar gas depend on angle from the Galactic Center since the the vector velocities of the LIC (away from the Galactic Center) and the G cloud (towards the Galactic Center) are different. The results for temperature (Figure~\ref{TandTurbvsGC} (left) and turbulent velocity (Figure~\ref{TandTurbvsGC} (right) show a wide scatter but no significant trends with angle from the Galactic Center direction.

The upwind direction of interstellar gas flowing into the heliosphere is the vector sum of the interstellar gas flow in the local standard of rest and the Sun's motion relative to the local standard of rest. The upwind velocity, temperature and direction have been studied using observations with the {\em EUVE, IBEX, Ulysses}, and {\em STEREO} spacecraft. A recent analysis of IBEX data by \cite{Swaczyna2018} gives $T=7691\pm230$~K, $v=26.21\pm0.37$ km~s$^{-1}$ and inflow Galactic coordinates ($l=3.5^{\circ}$, $b=+15.2^{\circ}$). The temperatures and turbulent velocities also show no trend with angle $\theta$ from the Galactic Center or the upwind direction.

\begin{figure*}[htb!]
\includegraphics[width=9.2cm]{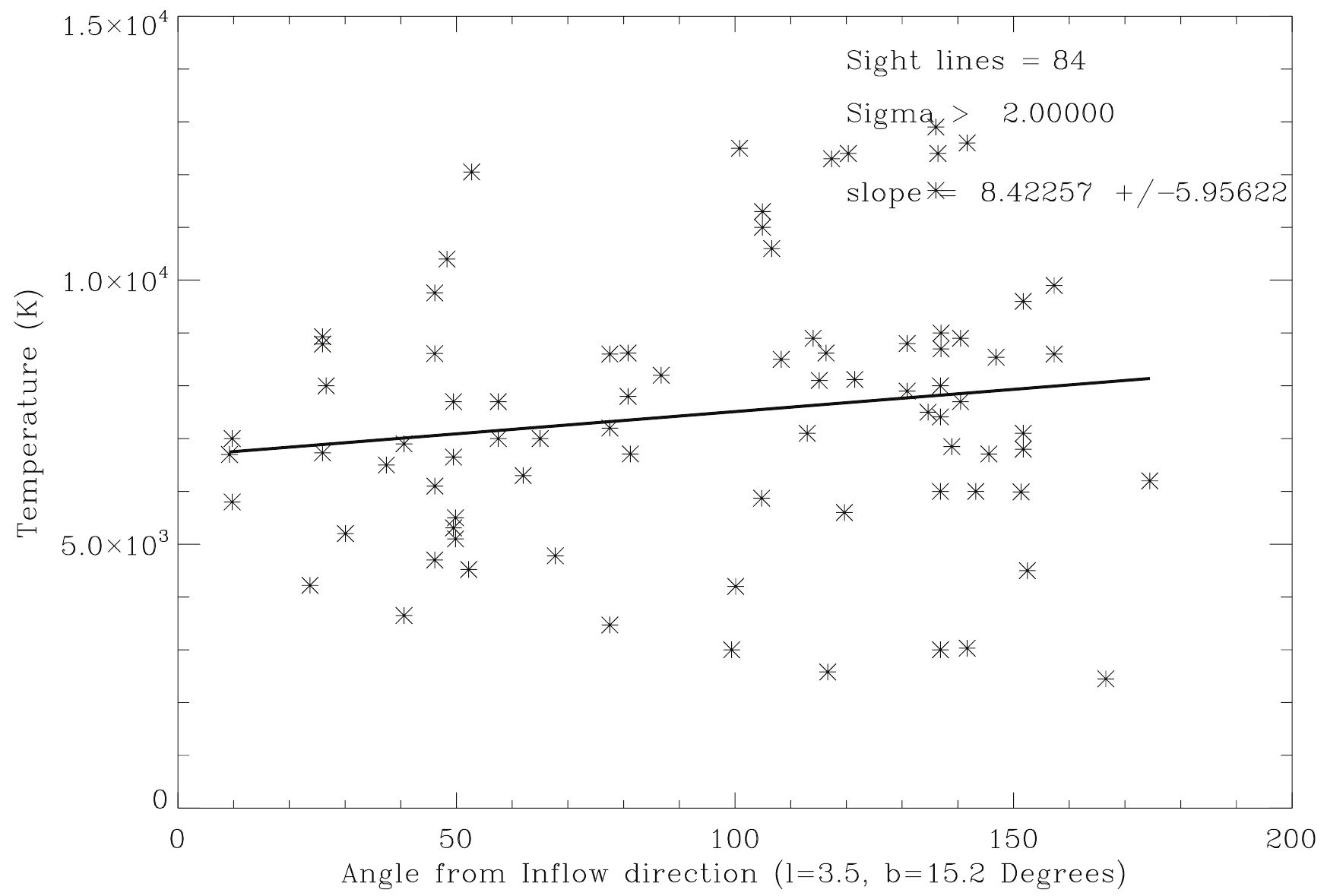}
\includegraphics[width=8.8cm]{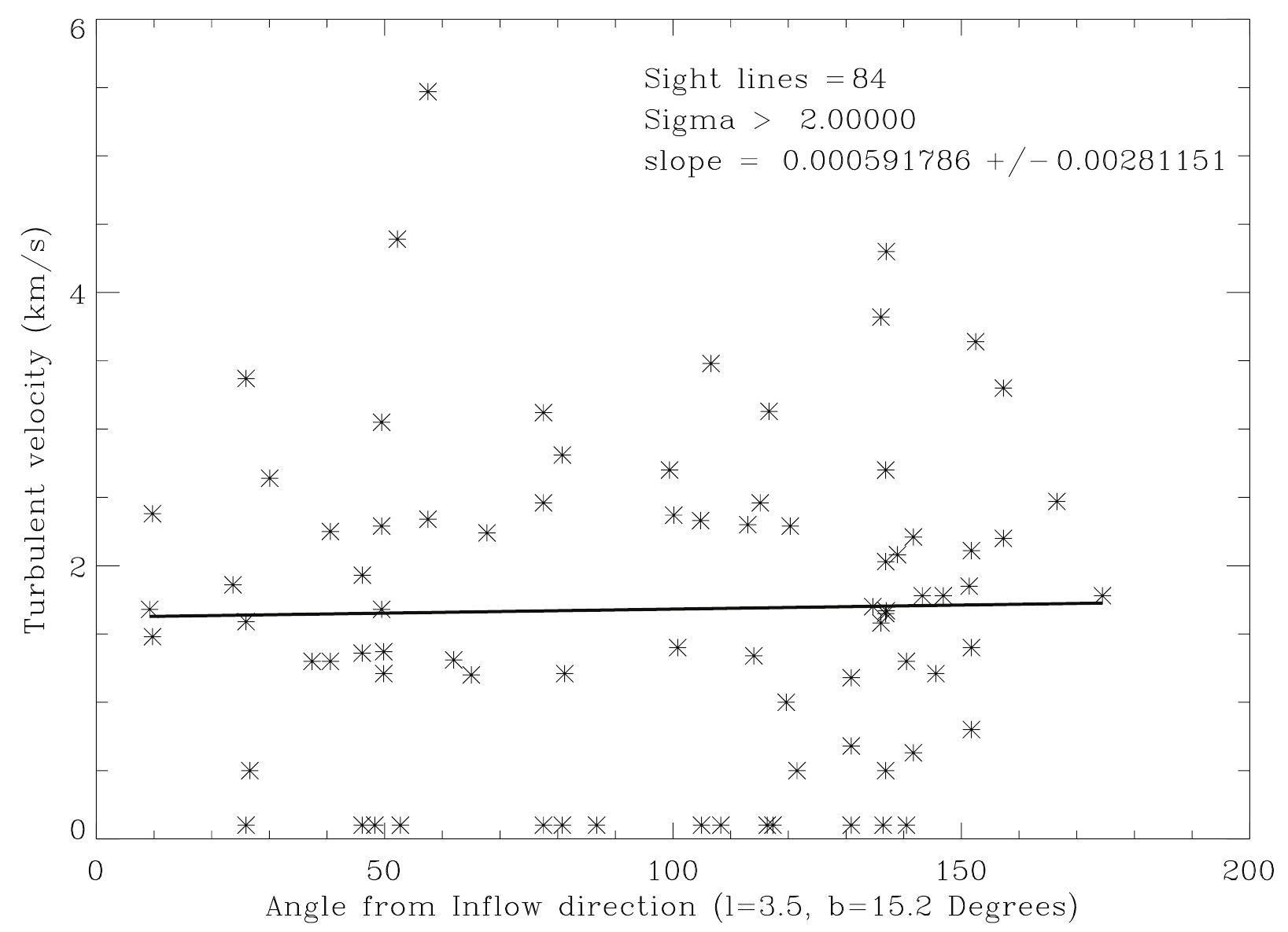}
\caption{Plots of temperatures ({\bf left}) and turbulent velocities ({\bf right}) vs. angle from the He~I inflow direction ($l=3^{\circ}.5$, $b=15^{\circ}.2$).\label{TandTurbvsinflow}}
\end{figure*}

\subsection{Do temperatures and turbulent velocities depend on the angle from the main ionizing source?}

\begin{figure*}[htb!]
\includegraphics[width=9.0cm]{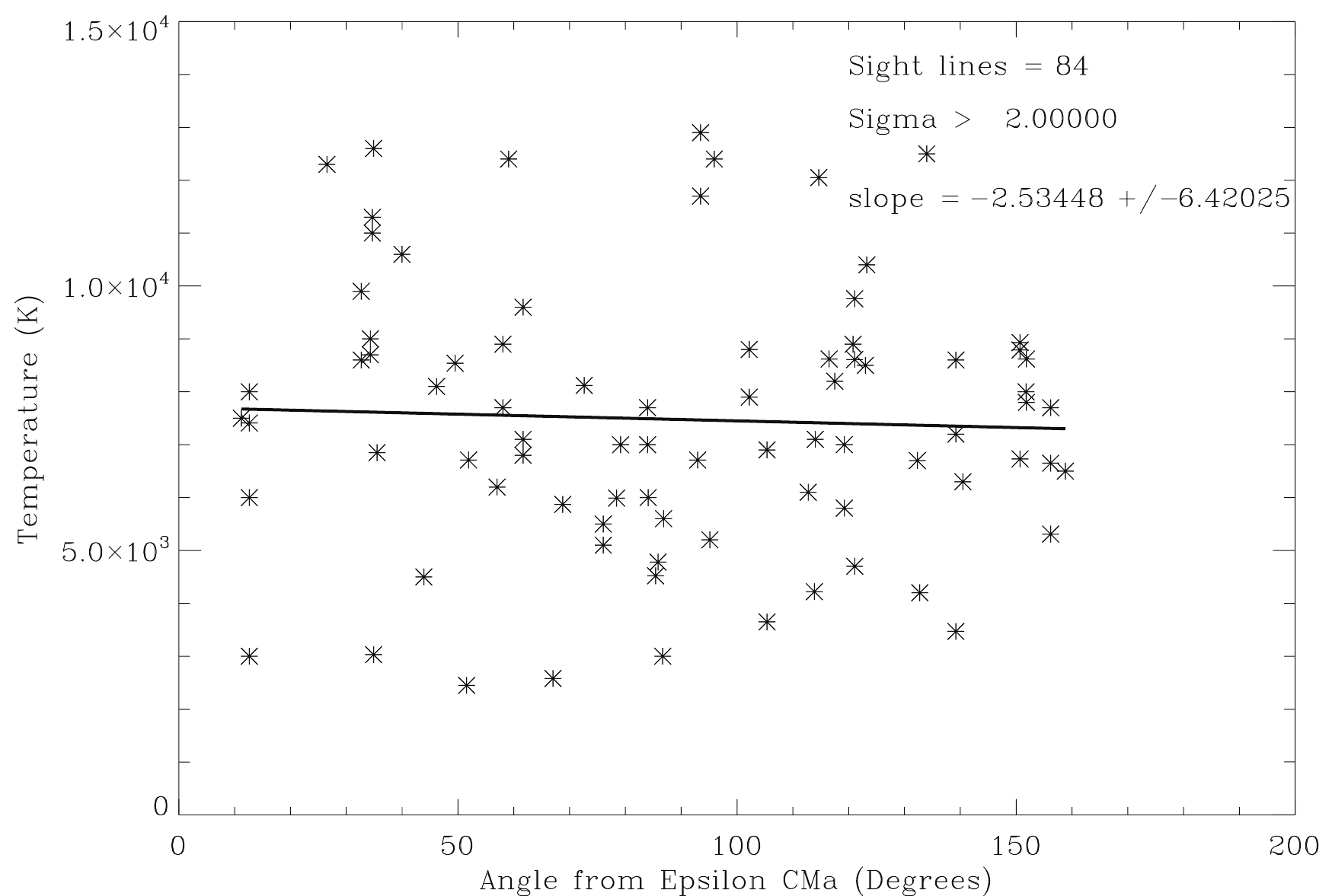}
\includegraphics[width=9.0cm]{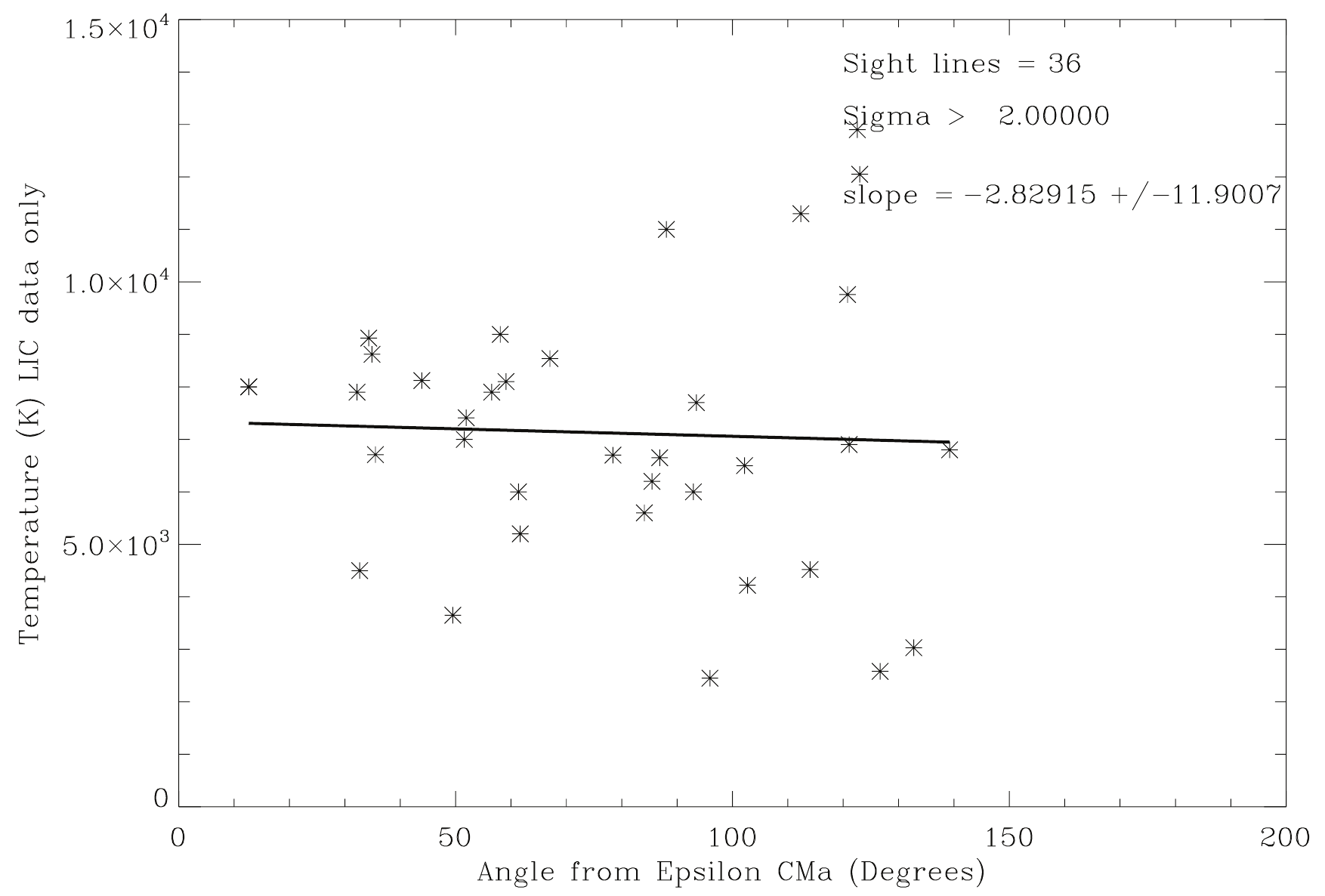}
\caption{Plots of sight line temperatures for the full data set ({\bf left}) and the LIC data set ({\bf right}) vs angle from $\epsilon$~CMa. The solid lines are least-squares linear fits to the data.  \label{TandTLICvsECMa}}
\end{figure*}

\begin{figure*}[htb!]
\includegraphics[width=9.0cm]{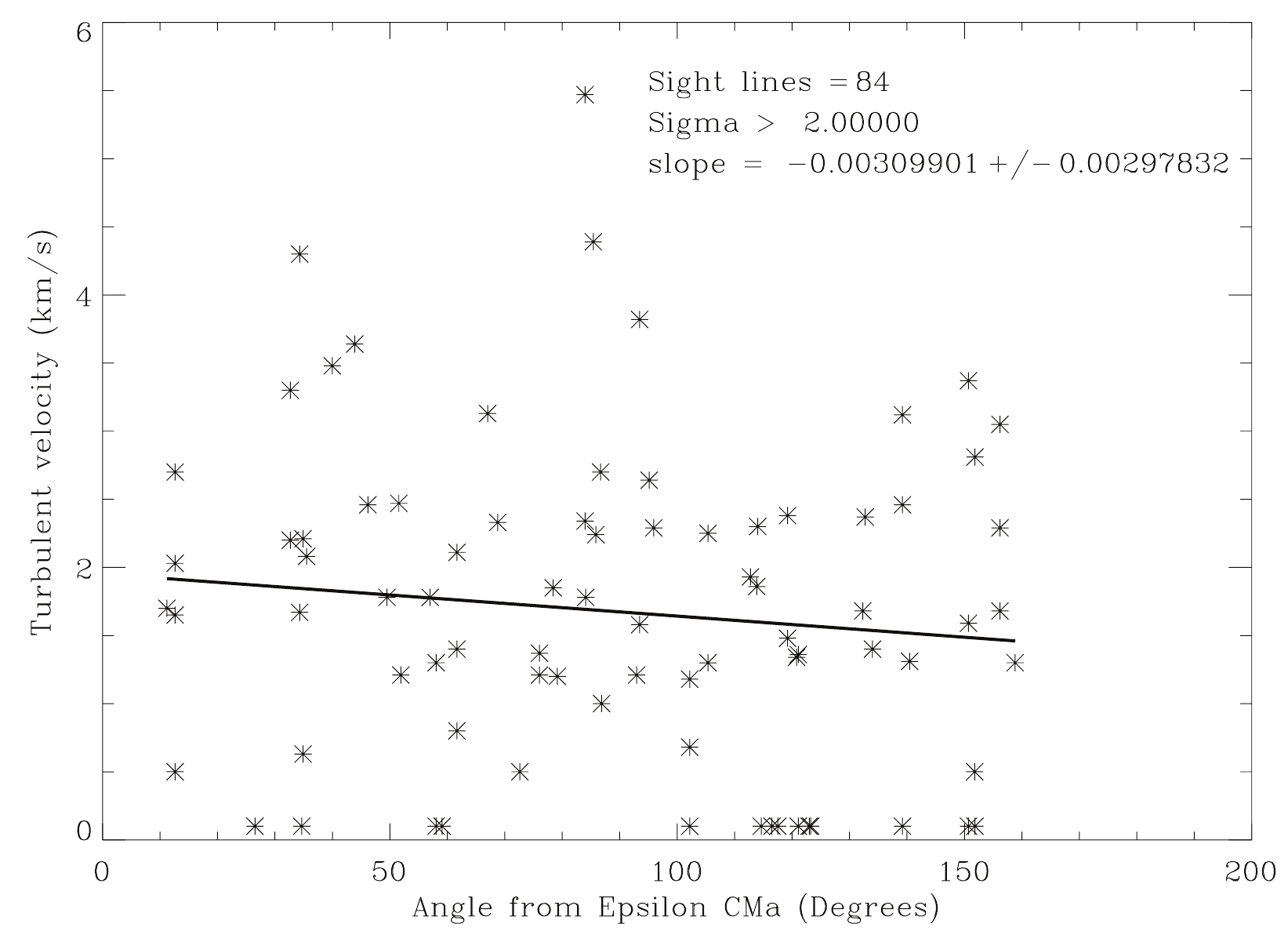}
\includegraphics[width=9.0cm]{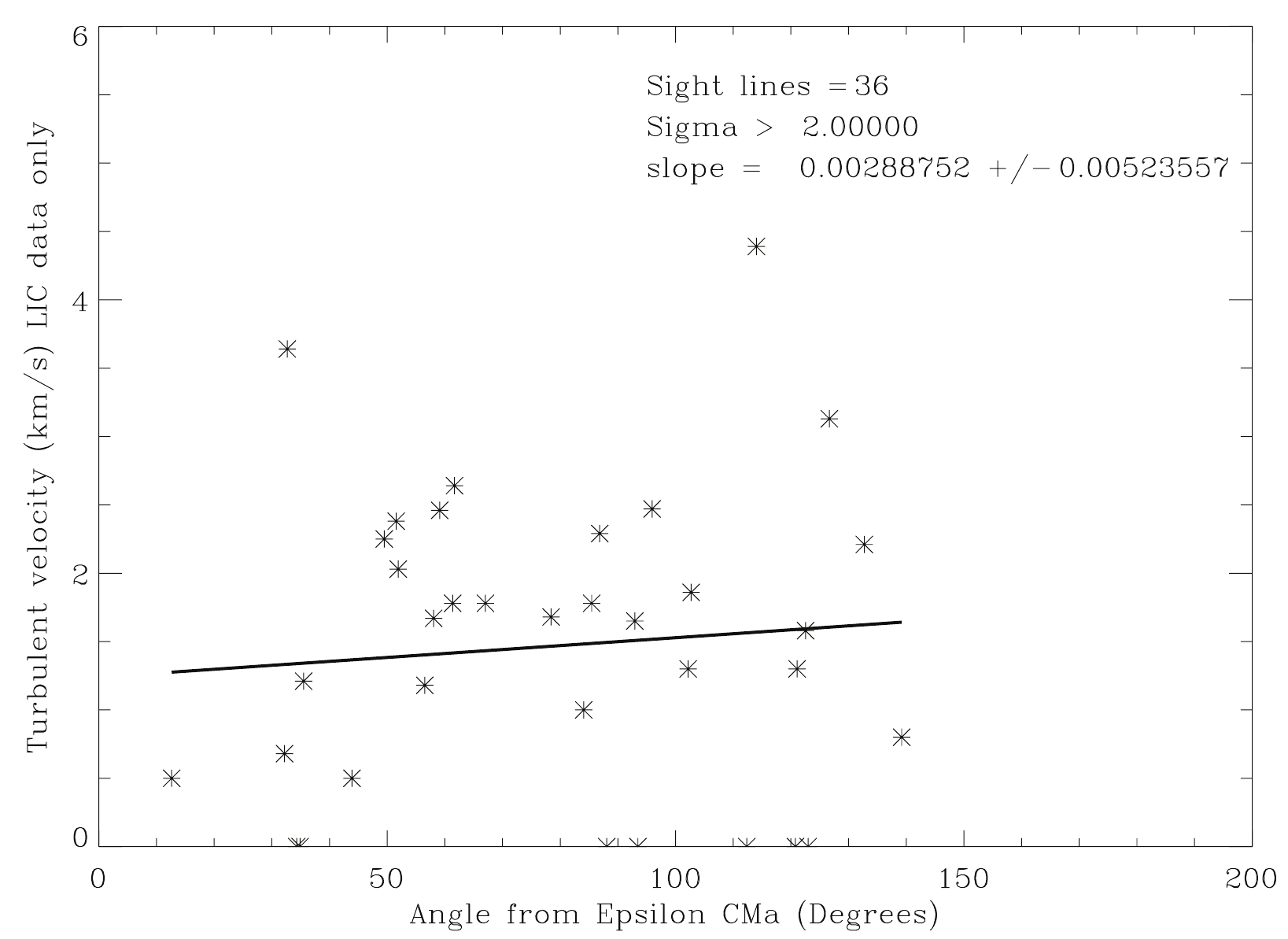}
\caption{Plots of sight line turbulent velocities for the full data set ({\bf left}) and the LIC data set ({\bf right}) vs angle from $\epsilon$~CMa. The solid lines are least-squares linear fits to the data. \label{TurbandTurbLICvsECMA}}
\end{figure*}

The main source of EUV radiation that ionizes hydrogen ($\lambda<912$\AA) is the star $\epsilon$~CMa (Galactic longitude $l=239.8^{\circ}$ and Galactic latitude $b=-11.3^{\circ}$) 
\citep{Vallerga1995}. Figure~\ref{TandTLICvsECMa} (left) shows the dependence of temperature and on angle from $\epsilon$~CMa for the full data set and  Figure~\ref{TandTLICvsECMa} (right) shows the dependence for the LIC data set. One might expect that the temperatures in the direction of $\epsilon$~CMa would be enhanced as photoionization of hydrogen produces free electrons with the kinetic energy of EUV photons minus the ionization energy of hydrogen. We see no such effect for the temperatures or turbulent velocities data with respect to the angle from $\epsilon$~CMa in Figure~\ref{TurbandTurbLICvsECMA}.

\subsection{Do temperatures and turbulent velocities depend on whether sight lines pass through the core or edge of the LIC?}

\begin{figure*}[htb!]
\includegraphics[width=9.2cm]{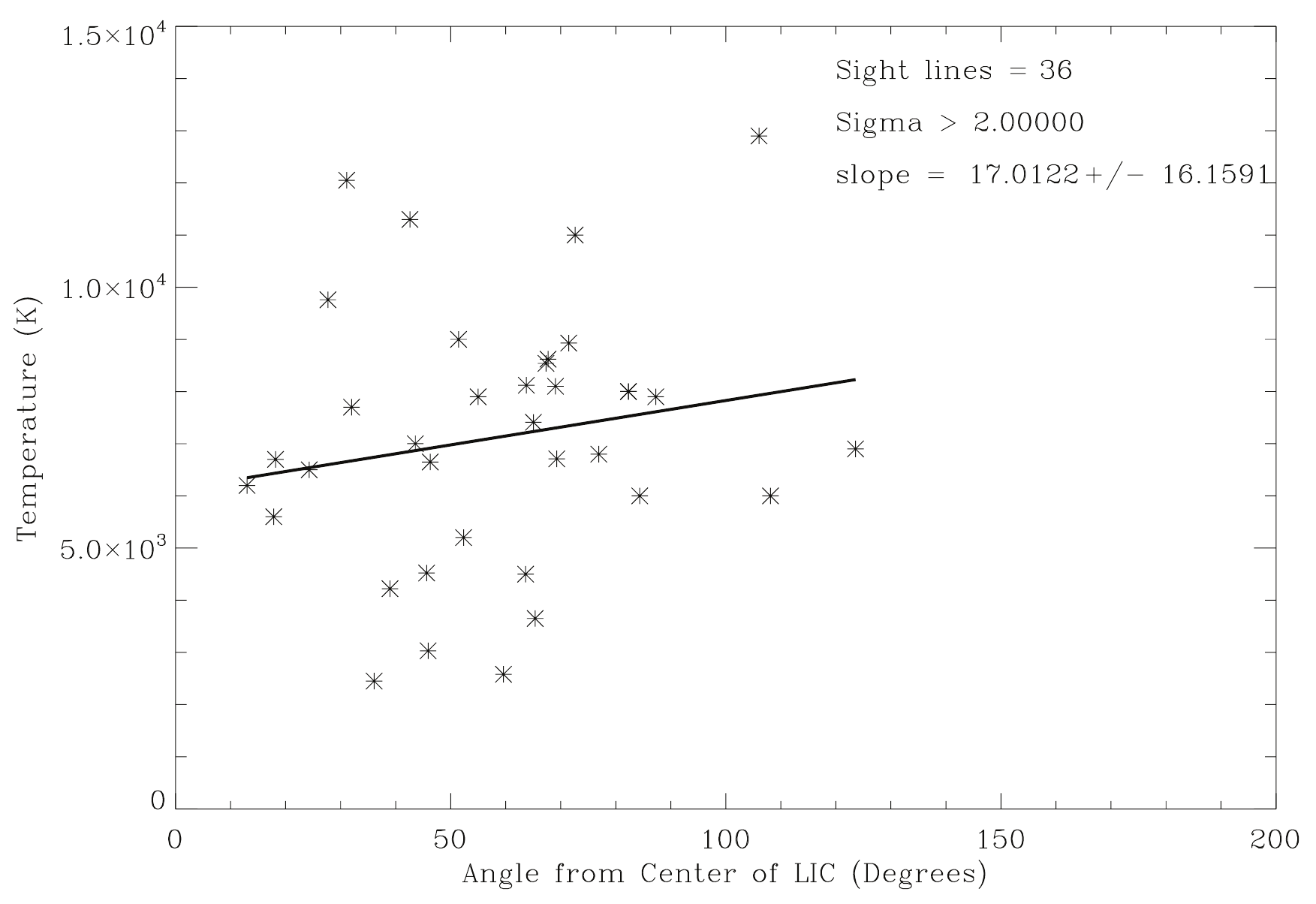}
\includegraphics[width=8.8cm]{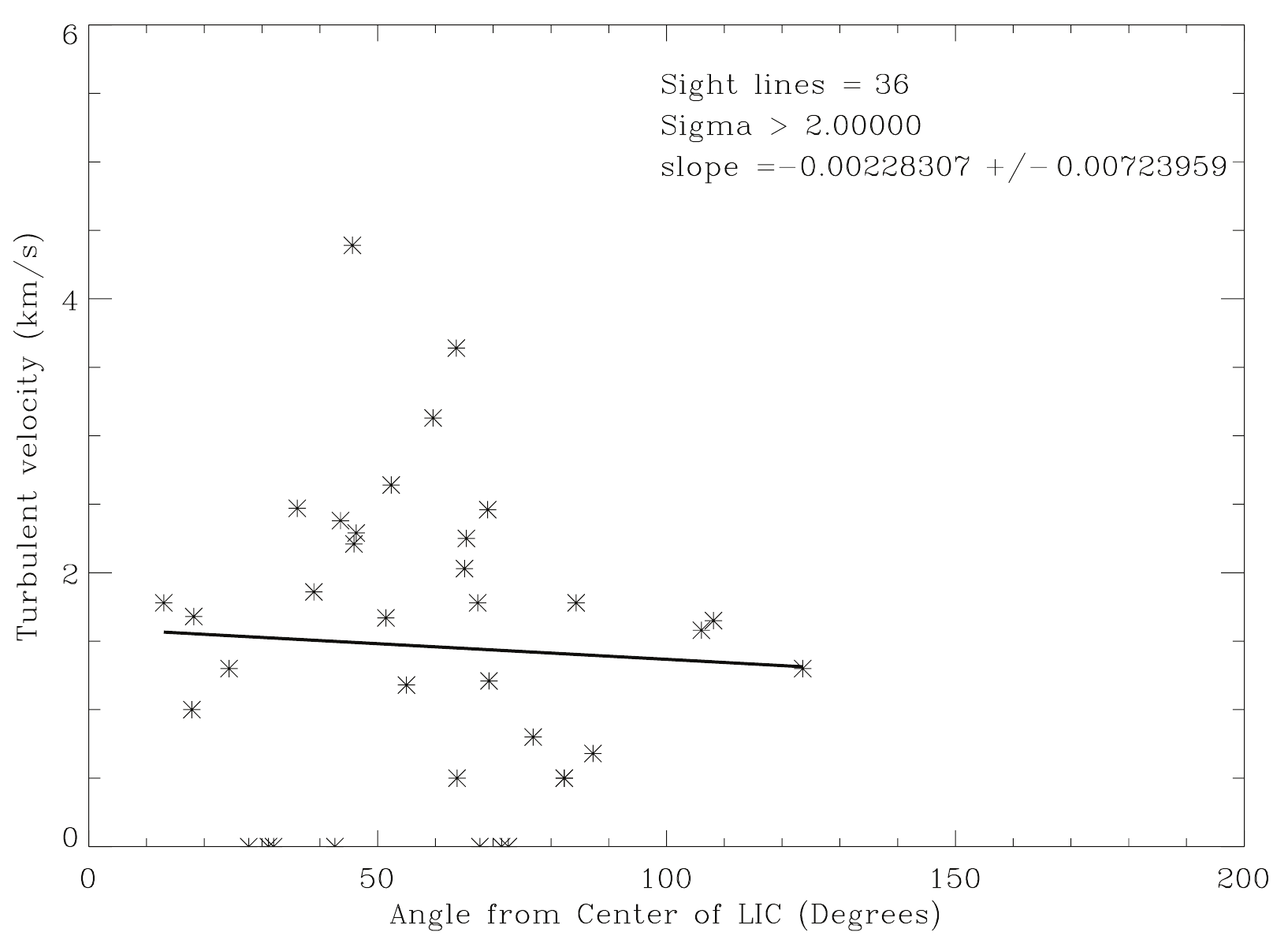}
\caption{Plots of sight line temperatures ({\bf left}) and turbulent velocities ({\bf right}) vs angle from the center of the LIC The solid lines are least-squares linear fits to the data. \label{TandTurbvsLICangle}}
\end{figure*}

Figure~\ref{TandTurbvsLICangle} addresses the question of whether the temperatures and turbulent velocities depend on whether sight lines penetrate through the core of the LIC (approximate coordinates $l=145^{\circ}$ and $b=0^{\circ}$) or its edge. There are no significant trends in the temperature or turbulent velocity on direction of sight lines through the LIC, indicating that self-shielding from external EUV radiation sources is not important for determining the temperature of the LIC plasma..

\subsection{Do temperatures and turbulent velocities depend on the neutral hydrogen column density?}

\begin{figure*}
\includegraphics[width=9.0cm]{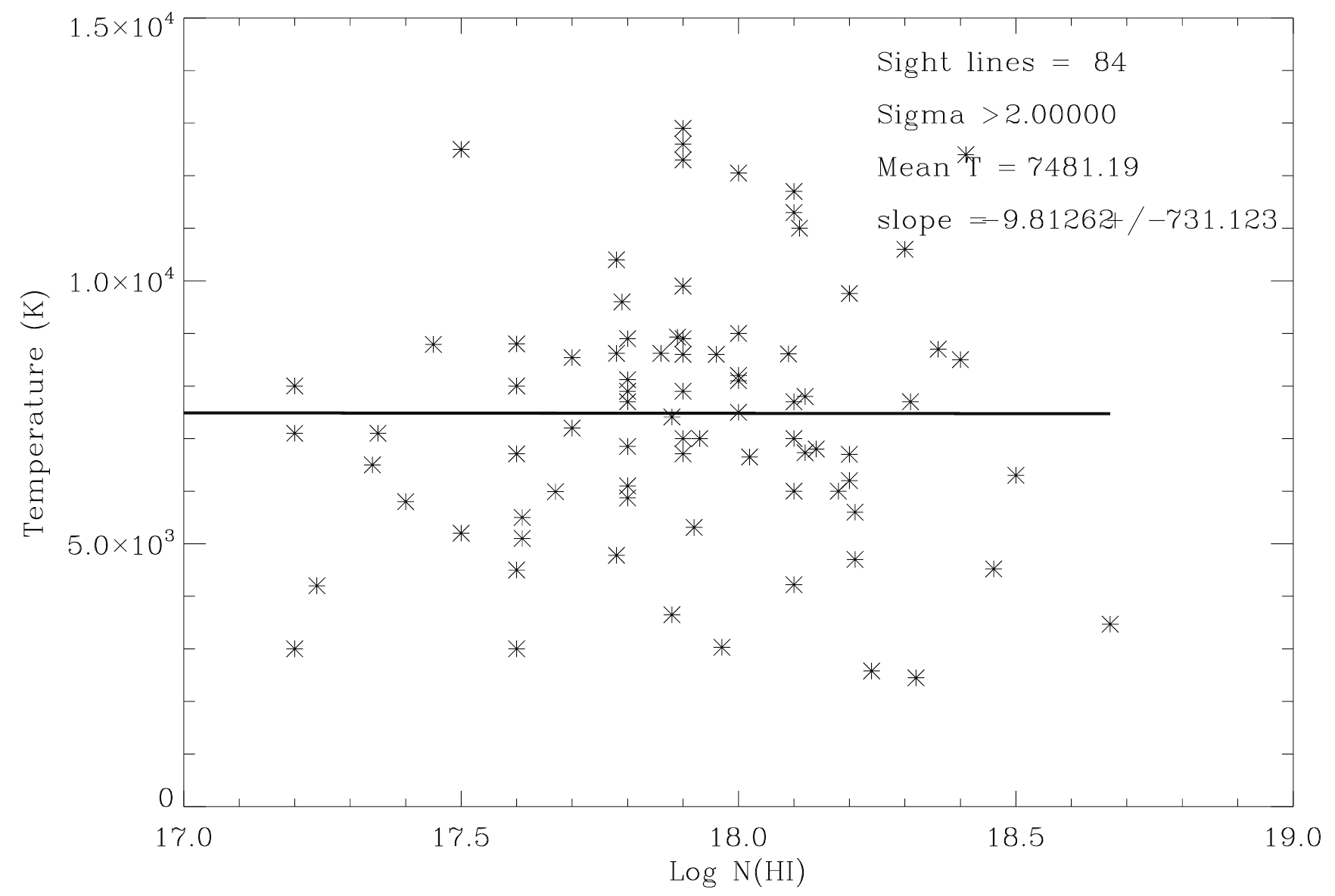}
\includegraphics[width=9.0cm]{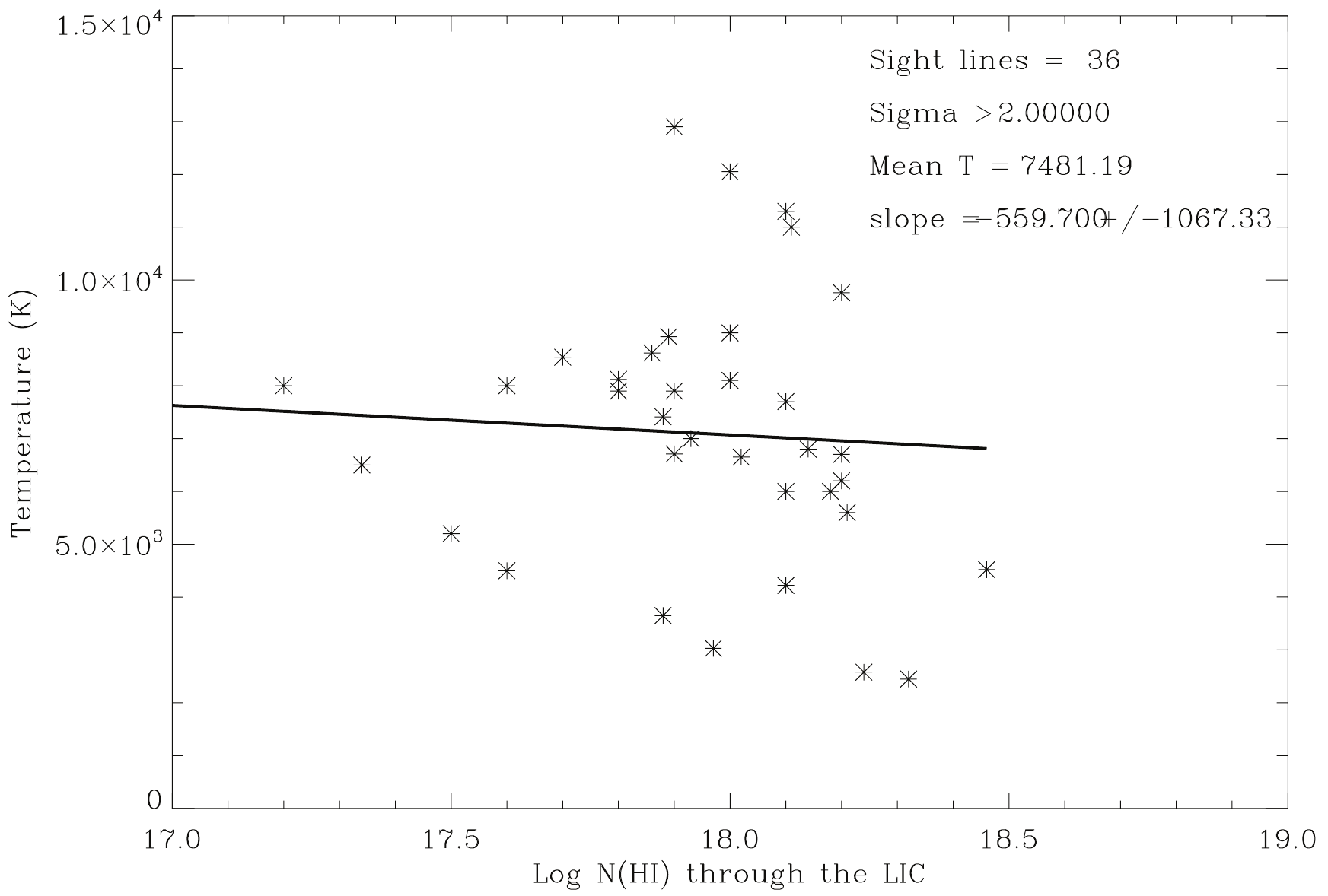}
\caption{Plots of temperatures for the full data set ({\bf left} and the LIC data set ({\bf right}) vs. the log of the neutral hydrogen column density. 
The solid lines are least-squares linear fits to the data. \label{TandTLICvslogNH}}
\end{figure*}

\begin{figure*}
\includegraphics[width=9.0cm]{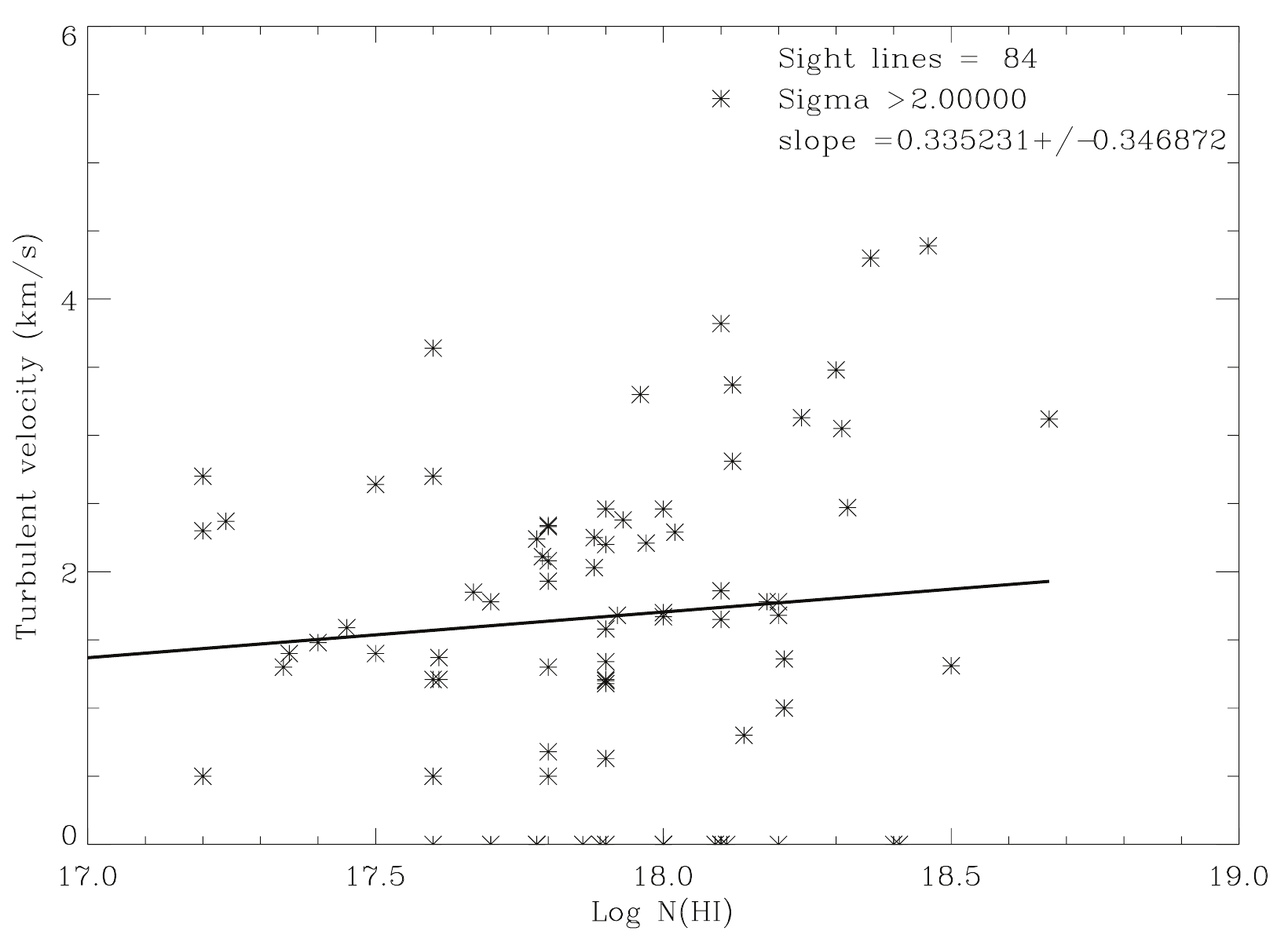}
\includegraphics[width=9.0cm]{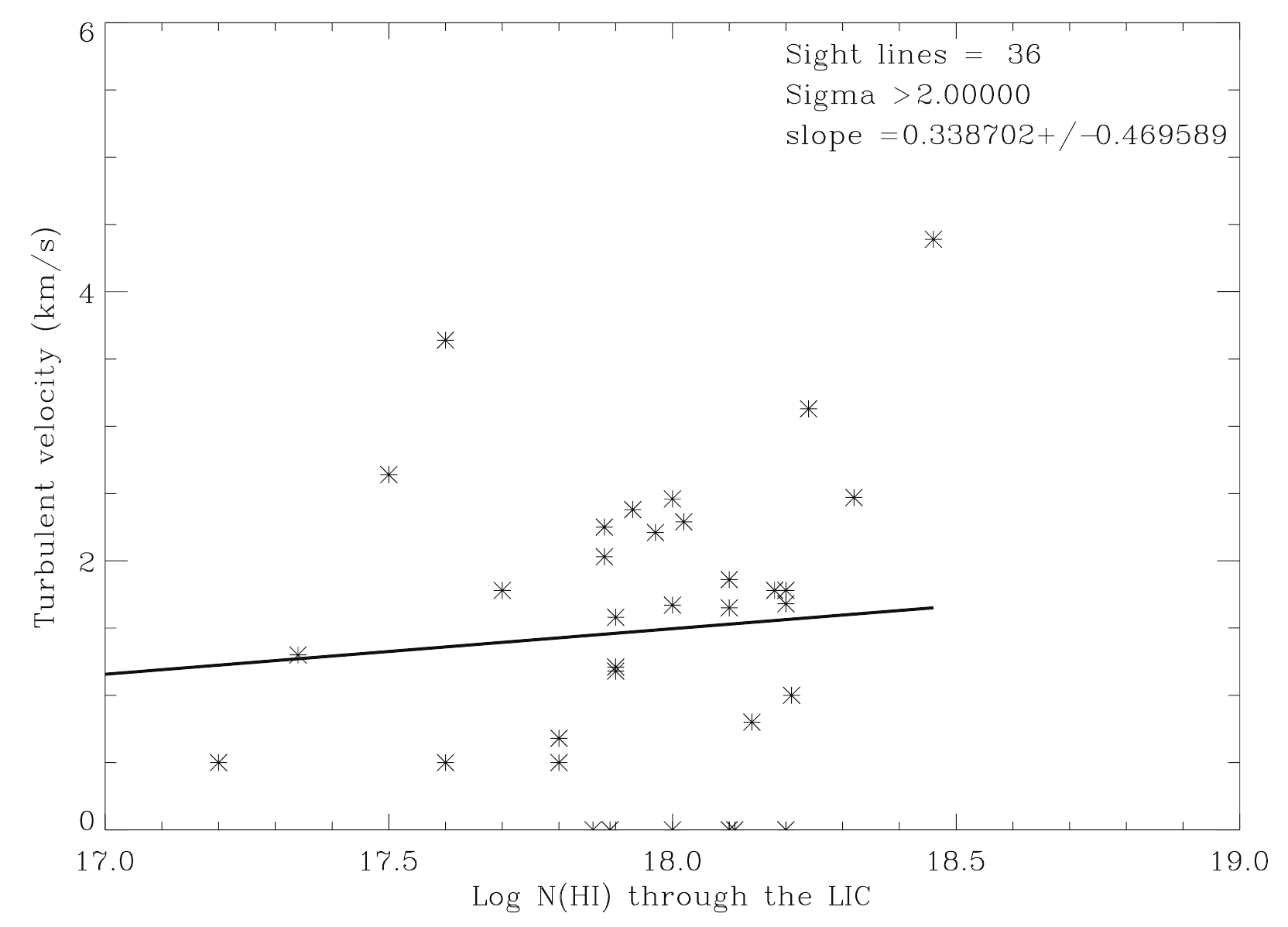}
\caption{Plots of turbulent velocities for the full data set ({\bf left}) and the LIC data set ({\bf right}) vs. the log of the neutral hydrogen column density. 
The solid lines are least-squares linear fits to the data. \label{TurbandTLICvslogNHLIC}}
\end{figure*}

We show plots of the temperature (Figure~\ref{TandTLICvslogNH} and turbulent velocity (Figure~\ref{TurbandTLICvslogNHLIC})
 vs $N$(H~I) for the full data set and the LIC data set. There are no significant trends of either the temperature or turbulent velocities on $N$(H~I). One might expect higher temperatures in low $N$(H~I) sight lines that would be in closer proximity to the external EUV radiation field or perhaps hot gas surrounding the clouds, but there is no evidence that either scenario is the case. 

\subsection{Do all velocity components seen in the Mg~II and Fe~II lines correspond to warm clouds?}

     Line of sight measurements of interstellar absorption provide information on the interstellar properties of gas in front of a star, but not where the cloud's absorbing gas begins and ends along the sight line. For stars within about 10~pc of the Sun, the detection of Lyman-$\alpha$ absorption blue shifted relative to the interstellar hydrogen absorption indicates the presence of interstellar neutral hydrogen that has change exchanged with energetic stellar wind protons \citep{Wood2005a}. This shows that the star is embedded in an interstellar cloud containing neutral hydrogen. The absence of this blue shifted astrospheric absorption could be explained by the star being embedded in fully ionized interstellar gas or the sight line being near the downwind direction. Table~6 lists the stars for which astrospheric absorption has been detected or definitely not detected. In the 15 cases where there are detections of blue shifted astrospheric absorption, the star must be located within the cloud identified in the sight line. In three cases where there are more than one cloud, there is uncertainty as to which of the 2 or 3 clouds envelopes the star but one of them must. For the eight  cases where no astrospheric absorption has been reported, it is likely that no cloud extends as far as the star.
  
     The one-dimensional extent of a cloud along a given line of sight is more complicated. Since the Sun is embedded in the outer region of the Local Interstellar Cloud (LIC), other clouds detected in the same sight line must lie beyond the LIC. It is likely that the G cloud also lies in front of other clouds as most of the sight line to the nearest star $\alpha$~Cen is in the G cloud with no detected absorption at the LIC cloud's velocity. If one assumes that $n$(H~I) in a cloud is the same as in the LIC (0.20 cm$^{-3}$) \citep{Slavin2008}, then a cloud's path length along the sightline is given by the ratio $N$(H~I)/0.20. While neutral hydrogen column densities $N$(H~I) can be measured with modest precision, there are no accurate methods for measuring $n$(H~I) in clouds other than the LIC.      

     High-resolution spectra of the Mg~II and Fe~II interstellar absorption lines observed twards nearby stars, primarily M dwarfs, provide a tool for identifying the fractional coverage of a cloud along the sight line to the star. Table~6 lists all of the stars within 10~pc of the Sun with measured $N$(H~I) and the number of clouds identified in each sight line. For each sight line, we identify the clouds located in front of the star by matching two criteria: (1)  that the measured interstellar radial velocities lie within 2 km~s$^{-1}$ of the velocities predicted from the cloud's velocity vector, and (2) the star lies within the cloud's morphological outline \citep{Redfield2008}. Cloud names are placed in parentheses when the star is located at the edge of or just beyond the cloud edge. 
     
     The cloud identification process is somewhat subjective as the cloud morphologies are not precisely known given the limited number of stars that were used to construct these morphologies. However, in all but three of the 52 velocity components in the 37 sight lines to stars within 10 pc, the number of measured interstellar velocity components is the same as the number of previously known clouds identified by the two-step procedure just described. For the sightline to GJ~15A one absorption component was observed but two clouds meet both selection criteria. This could result from both clouds being in the sight line or errors in the cloud outlines as the star is located at the edge of both the LIC and Hyades clouds. The agreement between the number of interstellar velocity components and the number of identified clouds for 14 of the 15 new sight lines that were not available when the cloud outlines were constructed by \cite{Redfield2008} provides strong support for the cloud outlines obtained at that time. The identification of a cloud near its edge is a strong determination that the cloud is actually in the sight line, since there is no alternative cloud that meets both criteria of measured radial velocity and fitting within the cloud boundary. We conclude that nearly all velocity components identified so far correspond to previously identified warm clouds.  
    
\subsection{Are the warm clouds in the CLIC closely packed or widely separated?} 

\begin{figure}[htb!]
\plotone{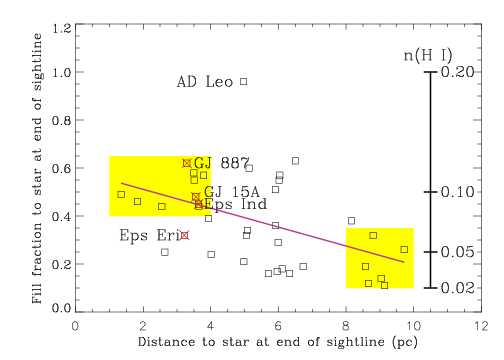}
\caption{Plot of the filling fraction of neutral hydrogen clouds along sight lines to stars vs stellar distance. On the righthand side is the scale of neutral hydrogen density if the sight lines are completely filled with the clouds. The upper colored area identifies sight lines that would be filled by clouds with a mean density $n$(H~I)$\approx0.10$~cm$^{-3}$. The lower colored area identifies sight lines with distant clouds that are widely separated. The 4 stars within 4~pc of the Sun with only LIC absorption and astrospheric absorption are marked by a red x. The identified stars are discussed in the text.\label{fillvsdist}}
\end{figure}

     The only cloud in the CLIC for which the neutral hydrogen density has been measured is the LIC, or more precisely the outer edge of the LIC where the heliosphere resides. Measurements of the inflowing neutral helium consistently predict $n$(H~I)$\approx 0.20$~cm$^{-3}$. \cite{Slavin2008} developed theoretical models for the LIC just outside of the heliosphere consistent with this neutral hydrogen density and other empirical constraints. For other clouds in the CLIC, there are measurements of the neutral hydrogen column density $N$(H~I) but not the number density $n$(H~I). It is tempting to assume that $n$(H~I) is roughly the same in the other CLIC clouds. If so, then it is simple to determine the extension of a cloud along a sight line from the measured $N$(H~I) assuming $n$(H~I)=0.20~cm$^{-3}$. 
     
    We can now test this assumption with data for 37 stars within 10~pc of the Sun. Table~6 lists the neutral hydrogen column densities for each cloud detected in the sight lines to all stars within 10~pc of the Sun that have measured $N$(H~I). Also listed are the filling fractions of the clouds along each sightline determined from the cloud extensions (assuming that $n$(H~I)=0.20~cm$^{-3}$) divided by the distance to the star. For sight lines with multiple clouds, the filling fraction is the sum for all of the clouds. Figure~\ref{fillvsdist} plots the filling fractions for the 37 sight lines. There are several interesting results shown in this figure:

\begin{itemize}
\item With the exception of AD~Leo (see Section 3.11), all of the sight lines have filling fractions smaller than 0.63. This maximum filling factor could be explained either by a sparcely filled CLIC with highly ionized gas located between the clouds observed along the sight line, or with clouds that entirely fill the sight lines but with smaller densities than the LIC value of 
$n$(H~I)=0.20~cm$^{-3}$, or by a combination of the two possibilities. The large sample of 37 relatively short sight lines with 53 absorption components and the detection of 15 astrospheres provides an opportunity to distinguish between these possibilities.
 
\item Sight lines to 11 of the 13 stars within 4~pc of the Sun have filling factors in the range 0.39 to 0.62. Also, an additional 5 more distant stars have filling factors that lie within this range. This result could be explained by the clouds having typical densities of $n$(H~I)$\approx0.10$~cm$^{-3}$ that fill the sight lines. The alternative explanation of sight lines to a large number of stars all being about 50\% filled seems contrived and less likely. The short sight line with the smallest fill factor (0.25) is towards Sirius AB (2.64~pc). This is not surprising because Sirius~B is a hot white dwarf with a surrounding H~II region of fully ionized hydrogen.

\item There are four stars within 4~pc of the Sun ($\epsilon$~Eri,  GJ~887, GJ~15A, and $\epsilon$~Ind, ) that have only LIC absorption in their sight lines and also astrospheric H~I Lyman-$\alpha$ absorption that requires that these stars are located inside of a cloud containing neutral hydrogen. The simplest model for these sight lines is that the LIC extends from the heliosphere to the astrosphere of each star and perhaps beyond. In this case, the mean filling factor toward these four stars is 0.47 and the mean neutral hydrogen density in the LIC is 0.094~cm$^{-3}$. The density n(H~I)=0.20~cm$^{-3}$ in the immediate environment of the Sun, therefore, appears to be unrepresentative of the mean neutral hydrogen density in the LIC.

\item Other clouds in the CLIC may have similar densities to $n$(H~I) $\approx0.10$~cm$^{-3}$ for the LIC. For example, the sight line to the nearest stellar binary system $\alpha$~Cen AB shows no evidence for LIC absorption, only G cloud absorption and astrospheric absorption likely produced by neutral hydrogen in the G cloud. After a minimal amount of LIC gas, the complete sight line to $\alpha$~Cen could be filled with G cloud gas with $n$(H~I)=0.10~cm$^{-3}$. We conclude that within 4~pc of the Sun, the clouds appear to be tightly packed, and that typical densities are $n$(H~I)$\approx 0.10$~cm$^{-3}$. However, a "Swiss cheeze" model with high density gas separated by voids cannot be ruled out by the available data.

\item Beyond 4~pc, there is a wider range of filling fractions with a clear trend of decreasing filling fraction with increasing distance as shown by the least-squares linear fit shown in Figure~\ref{fillvsdist}. For the sight lines between 4~pc and 7~pc the filling fractions have a wide range (0.16--0.63), but the seven sight lines in the range 8--10~pc all have filling fractions less than 0.38, and four of the sight lines have filling fractions of about 0.15. \cite{Gry2017} found that the warm gas filling factor in the sight line to $\alpha$~Leo (23.8~pc) is about 0.13. This trend of decreasing filling fraction with distance suggests that the CLIC clouds are becoming more widely separated with distance from the Sun, as was suggested by \cite{Redfield2008}, and that the inter-cloud gas with fully ionized hydrogen is occupying a larger fraction of sight lines to the more distant stars. The inter-cloud gas may be the same as the gas that fills most of the Local Cavity \citep{Linsky2019}.

\end{itemize}

\subsection{Is there a shock in the sight line to AD Leo?}

The short (4.97~pc) sight line to AD Leo is interesting for several reasons. The Galactic coordinates of AD~Leo ($l=216^{\circ}.5, b=+54^{\circ}.6$) place it at the edge of the LIC, Leo and NGP clouds \citep{Redfield2008}, but its interstellar radial velocity of 13.13~km~s$^{-1}$ is inconsistent with the predicted radial velocities of the LIC (7.23~km~s$^{-1}$), Leo (9.21 ~km~s$^{-1}$)  and NGP (16.22~km~s$^{-1}$) clouds, using the LISM Dynamical Model Kinematic Calculator\footnote{http://lism.wesleyan.edu/LISMdynamics.html}. As a result, AD Leo is not assigned to any known cloud and the high H~I column density may result from the sightline passing through a shock. Since the LIC and Leo clouds have essentially the same velocity amplitudes, the shock would not be from these two clouds. Instead, the NGP cloud has a velocity vector amplitude 13 km~s$^{-1}$ different from the LIC and Leo clouds and thus could be responsible for the shock. 

\cite{Gry2017} found interstellar absorption toward the star Regulus ($\alpha$~Leo) at $8.8\pm 0.2$~ km~s$^{-1}$, which they assigned to the LIC, and a second component at $14.4\pm 0.1$~km~s$^{-1}$. Since Regulus and AD~Leo are separated by only about $10^{\circ}$ and the radial velocity of the second component is similar to that seen toward AD~Leo, it is likely that these two components are formed in the same unassigned structure in front of AD~Leo and thus closer than 4.97~pc. Although the Local Leo Cold Cloud (LLCC) \citep{Peek2011} is in the same direction as Regulus (and AD~Leo) with a similar radial velocity (9.3 km~s$^{-1}$), \cite{Gry2017} argue that the LLCC lies beyond Regulus at a distance of $33.5\pm 11.3$~pc. Thus the anomalously large value of $N$(H~I) in front of AD~Leo is not the LLCC, but some other perhaps very interesting feature. 

If the high value of $N$(H~I) in front of AD~Leo is produced in a shock between the NGP and either the LIC or Leo clouds, then the radial velocity of the shock should be intermediate between the NGP and the other two clouds. The halfway radial velocity between the NGP and LIC is 11.7~km~s$^{-1}$ and between the NGP and Leo clouds is 12.7~km~s$^{-1}$. Both velocities are consistent with the observed interstellar velocity of 13.13~km~s$^{-1}$ towards AD~Leo. This spatial agreement between AD~Leo and the cloud interfaces and the velocity agreement provides evidence for a shock in the sight line to AD~Leo. This would be a second shock in the LISM, the first being the Cetus ripple discovered by \cite{Gry2014} located mostly in the southern Galactic hemisphere. 

\clearpage
\begin{deluxetable}{lcccccccccc}
\tablecolumns{11}
\tablewidth{0pt}
\tablecaption{Fill factors for sight lines to stars within 10 pc of the Sun} 
\tablehead{\colhead{HD} & \colhead{Star} & \colhead{$l$} & \colhead{$b$} & \colhead{d(pc)} & \colhead{Number}  & \colhead{logN(HI)} & \colhead{Filling} &  \colhead{Clouds} & 
\colhead{Astro-} & \colhead{Ref}\\ 
      & \colhead{Name} &    &        &          & \colhead{Clouds} &               & \colhead{fraction}  & \colhead{in front}  & \colhead{sphere}  & }
\startdata
 128620 & $\alpha$~ Cen ABC & 315.7 & --0.7 & 1.35 & 1 & 17.61 & 0.49 & G & Yes & 1\\
GJ 699  & Barnard's star & 31.0 & 14.1 & 1.83 & 1 & 17.72 & 0.46 & G & No & 3\\
95735 & GJ 411 & 185.1 & 65.4 & 2.55 & 1 & 17.84 & 0.44 & (LIC) & No & 3\\
48915 & SiriusAB & 227.2 & --8.9 & 2.64 & 2 & 17.4,17.2 & 0.25 &  (LIC), (Blue) & & 8\\
22049 & $\epsilon$~Eri & 195.8 & --48.1& 3.22 & 1 & 17.88 & 0.32 & LIC & Yes & 4\\
217987 & GJ 887   & 5.1 & --66.0 & 3.29 & 1 & 18.10 & 0.62 & LIC & Yes & 2\\
201091 & 61 Cyg A & 82.3 & --5.8 & 3.50 & 2 & 17.8,17.8 & 0.58 & Eri, Aql & Yes & 8\\
61421   & Procyon  & 213.7 & 13.0 & 3.51 & 2 & 17.9,17.6 & 0.55 & LIC, Aur & & 8\\
1326  & GJ 15A   & 116.7 & --18.4 & 3.56 & 1 & 18.02 & 0.48 &  (LIC), (Hya) & Yes & 2\\
209100 & $\epsilon$~Ind & 336.2 & --48.0 & 3.64 & 1 & 17.95 & 0.45 & (LIC) & Yes & 5\\
10700 & $\tau$~Cet & 173.1 & --73.4 & 3.65 & 1 & 18.01 & 0.44 & LIC & No & 1\\
 & GJ 273   & 212.3 & 10.4 & 3.79& 2 & 17.86,17.78 & 0.57 & LIC, Aur & & 2 \\ 
GJ 191 & Kapteyn's star & 250.5 & -36.0 & 3.93 & 1? & 17.98 & 0.39 & Blue & & 3\\
239960A & GJ 860A  & 104.7 & --0.0 & 4.01 & 1 & 17.78 & 0.24 &  (Eri) & Yes & 2 \\
 GJ 388 & AD Leo & 216.5 & 54.6 & 4.97 & 1 & 18.47 & 0.96 & (LIC), (Leo) & No & 1\\
26965 & 40 Eri A & 200.8 & --38.1 & 4.98 & 1 & 17.8 & 0.21 & LIC & No & 6\\
GJ 873. & EV Lac  & 100.6 & --13.1 & 5.05 & 1 & 17.97 & 0.32 & (Hyades) & Yes & 1\\
165341 & 70 Oph A & 29.9 & 11.4 & 5.08 & 3 & 17.8,17.1,17.5 & 0.34 & G, (Aql), (Mic) & Yes & 8\\
187642 & $\alpha$~Aql & 47.1& -8.9 & 5.13 & 3 & 17.9,17.9,17.5 & 0.60 & Aql, Eri, (Mic) & & 8\\ 
36395 & GJ 205  & 206.9 &--19.4 & 5.70 & 2 & 17.60,17.24 & 0.16 &  (LIC), unassigned & Yes & 2\\
 & GJ 754 & 352.4 & --23.9 & 5.91 & 2 & 18.16,17.62 & 0.51 & unassigned, (Aql) & & 9\\
 & GJ 588   & 332.7 & 12.1 & 5.92 & 1 & 18.12 & 0.36 & G & & 2\\
 155886 & 36 Oph A & 358.3 & 6.9 & 5.96 & 1 & 17.85 & 0.17 & G & Yes & 7\\
 GJ 285 & YZ CMi   & 215.9 & 13.5 & 5.99 & 2 & 17.89,17.45 & 0.29 & LIC, (Aur) & Yes & 2\\
191408 & GJ 783A & 5.2 & --30.9 & 6.02 & 1 & 18.31 & 0.55 &  (Mic) & & 11\\
20794 & 82 Eri & 250.7 & --56.7 & 6.04 & 1? & 18.33 & 0.57 & G & & 3\\ 
190248 & $\delta$ Pav & 329.8 & --32.4 & 6.10 & 2 & 17.82,17.55 & 0.18 & unassigned, (Vel) & & 9\\
79210 & GJ 338A  & 164.9 & 42.7 & 6.33 & 1 & 17.79 & 0.16 & LIC & Yes & 2\\
152751 & GJ 644B  & 11.0 & 21.1 & 6.50 & 1 & 18.40 & 0.63 & (Mic) & & 2\\
131156 & $\xi$~Boo A & 23.1 & 61.4 & 6.73 & 1 & 17.92 & 0.19 & Gem & Yes & 1\\
HIP86287 & GJ 686 & 42.2 & 24.3 & 8.16 & 1 & 18.28 & 0.38 & LIC & & 9\\
115617 & 61 Vir & 311.9 & 44.1 & 8.57 & 1 & 18.01 & 0.19 & (NGP) & Yes? & 11\\
39587 & $\chi^1$~Ori & 188.5 & --2.7 & 8.66 & 1 & 17.93 & 0.12 & LIC & No & 1\\
192310 & GJ 785 & 15.6 & --29.4 & 8.81 & 3 & 17.96,17.90,16.17 & 0.32 & (Vel),(Mic),(LIC)? & & 10\\
23249 & $\delta$~Eri & 198.1 & --46.0 & 9.04 & 1 &17.88 & 0.14 & LIC & Yes & 1\\
20630 & $\kappa^1$~Cet & 178.2 & --43.1 & 9.14 & 2 & 17.5,17.5 & 0.11 & LIC,(Hyades) & No & 8\\
197481 & AU Mic & 12.7 & --36.8 & 9.72 & 1 & 18.36 & 0.26 & (Mic) & No & 1\\
\enddata
\tablerefs{(1) \cite{Wood2005b}; (2) \cite{Wood2021}; (3) \cite{Youngblood2022}; (4) \cite{Dring1997};
(5) \cite{Wood1996}; (6) \cite{Wood1998}; (7) \cite{Wood2000}; (8) \cite{Redfield2008};
(9) \cite{Zachary2018}; (10) \cite{Edelman2019}; (11) This paper.}
 \end{deluxetable}

\section{Discussion}

\subsection{The Development of Empirical LISM studies}

Empirical studies of the local ISM can be viewed as having proceeded through three stages driven by the increasing availability of high-resolution ultraviolet spectra and the constraint of matching the properties of the gas flowing from the LISM into the heliosphere. The first stage consisted of studies of individual sight lines to nearby bright stars --- first by analyzing ground-based observations of the Ca II H and K lines, and then by analyzing ultraviolet spectra from the {\em Copernicus} satellite and the HRS instrument on {\em HST}. These studies provided measurements of radial velocities, temperatures, turbulent broadening, hydrogen and metal column densities, and electron densities of interstellar matter in these sight lines. 
Subsequently, \cite{Redfield2004b} measured the properties for 50 velocity components along the sight lines to 29 stars using high-resolution spectra from the STIS instrument on {\em HST}. These studies could not determine where along the sight line the absorption occurs or whether these properties are homogeneous or the mean of variations along the sight line. 

The recognition that the flow of interstellar gas is in the form of co-moving structures, now called clouds, is the second stage in the empirical study of the LISM. It began with the discovery by \cite{Crutcher1982} that radial velocities in many sight lines in directions away from the Galactic Center are consistent with a coherent flow from the direction of the Scorpio-Centaurus Association, and that the properties of the gas in this flow are consistent with the properties of neutral helium atoms flowing into the heliosphere from the LISM \citep{Witte1993}. The region from which this flow originates is now called the Local Interstellar Cloud (LIC). Subsequently, \cite{Lallement1992} found that the flow of gas observed for sight lines in the Galactic Center direction are consistent with a different flow vector that they named the Galactic cloud and is now called the G cloud. \cite{Lallement1994} noted that the sight line to Sirius~A shows a second velocity feature at $-5.7\pm 0.2$~km~s$^{-1}$ relative to the LIC absorption. A similar blue shifted absorption component in the direction of 
$\epsilon$~CMa observed by \cite{Gry2001} confirmed that this extra absorption is from a third cloud called the Blue cloud.

With the accumulated data base of 270 radial velocity measurements towards 157 stars within 100~pc of the Sun, \cite{Redfield2008} identified 15 velocity vectors. The validity of these clouds and their morphologies was confirmed by \cite{Malamut2014}, who found that nearly all of the newly observed velocity components that lie within the morphologies of clouds previously identified by \cite{Redfield2008} have radial velocities consistent with the cloud vector velocities. In this paper we find that the radial velocities of 49 out of 52 velocity components towards stars within 10~pc have radial velocities consistent with the clouds in their directions. Although the multi-cloud model fits essentially all of the available data, the assumption of discrete clouds with constant internal flows and finite edges may be unrealistic. An alternative model proposed by \cite{Gry2014} in which the LISM consists of one cloud with internal velocity gradients filling all of space within 9~pc may be more realistic, although it does not fit the data as well as the multi-component model \citep{Redfield2015}. Further support for the multi-component model comes from the scintillation of radio emission from point source quasars \citep{Linsky2008}.

The third stage in the development of empirical LISM studies beginning with this paper is the recognition that the temperatures and turbulent velocities within the LIC and other clouds are not homogeneous, but are distributed over a wide range. Whether or not this distribution is random remains to be seen. A future stage in understanding of the LISM would be the identification of the physical causes responsible for these variable parameters solely on the basis of observations. A related question is whether the clouds completely fill space within about 4~pc as discussed in Section 3.11, or whether an inter-cloud medium separates the clouds especially at larger distances. \cite{Breitschwerdt2000} proposed that the LIC and other warm clouds are produced by fragmentation due to hydromagnetic Rayleigh-Taylor instabilities that occur where the Local Bubble and Loop~I interact. This formation mechanism would be consistent with the clouds being isolated structures separated from other clouds by an ionized inter-cloud medium.

\subsection{The Development of Theoretical ISM Studies}

In parallel with the empirical studies, theoretical models and physically based  simulations have developed in two stages. The first stage involved models that included many heating and cooling processes, but assumed that there is energy and pressure balance in an assumed quiescent and static ISM. Theoretical models of a multi-component interstellar medium \citep{Field1969,Wolfire1995} assume that the balance between heating and cooling processes plays the critical role in identifying the likely temperature-pressure structures in the ISM. In these models, warm interstellar gas is usually modeled as two stable phases: the warm neutral medium (WNM) consisting of neutral hydrogen and other species, and the warm
ionized medium (WIM) in which hydrogen is fully ionized. The WNM and WIM can co-exit at the same thermal pressure given the model assumptions. The WIM is often called an H~II region or a Str\"omgren sphere surrounding a star emitting strong EUV radiation \citep[cf.][]{Linsky2019}.

Since there is strong evidence that near the heliosphere the LIC is partially ionized, n$_e$/n$_H=0.07$~cm$^{-3}$/0.195~cm$^{-3}$=0.35 \citep{Slavin2008}, neither of the two warm models may provide a useful prototype for the properties of the LIC and other nearby clouds. Our data on temperatures in the LIC and nearby partially ionized clouds can test these models.

In the second stage, numerical simulations of the interstellar medium powered by supernova explosions and winds and radiation from hot stars predict a very different ISM. The simulations of \cite{deAvillez2005} and \cite{deAvillez2012}, for example, show that dynamical phenomena resulting from shock waves and instabilities create a time-dependent non-equilibrium ISM in which there are no steady-state phases and all parameters (temperatures, turbulent velocities, densities, and magnetic fields) have a wide range of values both spatially and over time. Also, highly non-linear heating and cooling processes mean that time independent energy balance is not realistic. In these simulations more than half of the mass is in thermally unstable phases predicted by the steady state models. Are these simulations relevant for describing the LISM embedded inside an old supernova remnant, or are the steady-state theoretical models of a more quiescent LISM a better approximation? The diverse properties of the LIC and the clouds provides an opportunity to begin testing the very different models.

\subsection{Tests of Theoretical Models}

One of the predictions of the steady-state theoretical models is that temperatures in both the WNM and WIM lie in the thermally stable range 3,000--10,000~K depending on the rates of heating and cooling. Temperatures below 3,000~K are predicted to be unstable due to rapid cooling to very low temperatures becoming CNM gas. By contrast, the probability distribution function for temperatures is roughly flat between 100~K and $10^6$~K in the \cite{Breitschwerdt2021} simulations. We measured temperatures in a number of clouds with $T>10,000$~K, the highest temperature with small uncertainty being $T=12,050^{+820}_{-790}$~K for the sight line to V368~Cep through the LIC. This apparently high temperature limit is rough consistent with energy balance in the WNM and WIM and the rapid increase of Lyman-$\alpha$ cooling with increasing temperature.

Table~7 lists the sight lines through clouds with low interstellar gas temperatures. There are 13 velocity components in 10 sight lines with temperatures below 3,500~K, but many have large uncertainties. Four velocity components have temperatures including one $\sigma$ positive errors that lie below 3,500~K --- HD~72905 ($\pi^1$~UMa), HD~129333 (EK~Dra), and HD~220657 ($\upsilon$~Peg). These stars do not lie in similar directions, but both EK~Dra and $\upsilon$~Peg are at the edges of clouds. However the sight line to $\pi^1$~UMa passes close to the center of the LIC. These velocity components provide evidence for interstellar matter somewhat cooler than $T<3,000$~K, the nominal unstable temperature regime for WNM models. The nominal temperatures for the two components toward $\upsilon$~Peg are 1,000~K and 1,700~K, but the one $\sigma$ errors in these temperature measurements are large. The low temperatures are distributed among seven known clouds and two sight lines without known clouds. In most cases the sight lines pass through the edges or just outside of clouds. We note that the sight line to 70~Oph includes low temperatures through three clouds, and the sight line to $\upsilon$~Peg contains low temperatures through two clouds. These two stars are well separated in Galactic coordinates. However, the sight lines to SAO~28753, $\eta$~UMa, and EK~Dra are in similar directions, suggesting that these low temperatures could have a common origin. Since two of these identifications are unassigned to any known cloud and the sightline to the third (EK~Dra) passes through the edge of or just outside of the LIC, the low temperatures may occur in interstellar matter outside of partially ionized clouds, perhaps in the WNM.  The Local Leo Cold Cloud (LLCC) located between 11.3~pc and 24.3~pc from the Sun \citep{Peek2011} has a temperature of 15--30~K, which is definitely CNM. None of the low temperature sight lines listed in Table~7 is near the LLCC (centered at $l=222^{\circ}, b=44^{\circ}$).

\begin{table}
\caption{Low temperature sightlines} 
\begin{center}
\begin{tabular}{lccccccccc}
\hline\hline
HD & Name & $l$ & $b$ & d(pc) & $<v>$ & $T$ & $\xi$ & Ref & cloud \\ \hline
48915  & $\alpha$~CMa & 227.2 & -8.9 & 2.6 & 12.70 & $3000^{+2000}_{-1000}$ & $2.7\pm 0.3$ & 2 & Blue\\ 
GJ 873 & EV Lac             & 100.6 & --13.1 & 5.0   &  $4.44\pm0.43$ & $3030^{+670}_{-610}$ & $2.21^{+0.22}_{-0.23}$ & 1 & (Hyades??)\\
165341 & 70 Oph & 29.9 & 11.4 & 5.1 & $-26.50\pm0.07$ & $2700^{+3000}_{-2300}$ & $3.64^{+0.42}_{-0.44}$ & 2 & G \\
165341 & 70 Oph & 29.9 & 11.4 & 5.1 & $-32.53\pm1.30$ & $1700^{+2100}_{-1700}$ & $3.3\pm1.1$ & 2 & (Oph)\\
165341 & 70 Oph & 29.9 & 11.4 & 5.1 & $-43.34\pm0.92$ & $3300\pm2100$ & $2.31\pm0.37$ & 2 & (Aql)\\
 72905  & $\pi^1$~UMa  & 150.6 & 35.7 & 14.5 &  $12.91\pm0.43$ & $2450^{+740}_{-660}$ & $2.47^{+0.10}_{-0.11}$ & 1 & LIC\\
 43162  &  GJ 3389         & 230.9 & --18.5 & 16.7 & $17.1\pm0.43$ & $3200^{+2680}_{-2330}$ & $5.76^{+0.25}_{-0.26}$ & 1 & LIC\\
116956 & SAO 28753 & 113.7 & 59.5 & 21.7 & 7.7  & $2540^{+2450}_{-1830}$ & $1.66^{+1.21}_{-1.66}$ & 1 & unassigned\\ 
120315 & $\eta$ UMa  & 100.7 & 65.3  & 30.9 & $2.6\pm3.4$ & $0^{+4400}_{-0}$ & $5.6^{+0.9}_{-1.1}$ & 2 & unassigned\\
129333 & EK Dra & 105.5 & 49.0 & 34.6 & $-2.43\pm 0.32$ & $2580^{+280}_{-250}$ & $3.13^{+0.05}_{-0.05}$ & 1 & (LIC)\\
93497 & $\mu$~Vel & 283.0 & 08.6 & 35.9 & $-4.38\pm0.43$ & $3470^{+580}_{-590}$ & $3.12^{+0.11}_{-0.10}$ & 1 & G\\
220657 & $\upsilon$~Peg & 98.6 & --35.4 & 53.1 & $8.8\pm1.2$ & $1000^{+1900}_{-1000}$ & $3.46^{+0.61}_{-0.63}$ & 2 & (Hyades)\\
220657 & $\upsilon$~Peg & 98.6 & --35.4 & 53.1 & $1.73\pm0.39$ & $1700^{+1100}_{-900}$ & $3.93\pm0.22$ & 2 & (Eri)\\

\hline\hline
\end{tabular}
\end{center}
References: (1) This paper; (2) \cite{Redfield2004b}.\\
\end{table}

\subsection{Possible causes for the diversity of Cloud Temperatures} 

The observed diversity of temperatures inside the LIC and in other partially ionized clouds raises the question of what physical processes could be responsible. Localized heating by hydromagnetic shocks is an obvious candidate that needs to be pursued with high spatial resolution observations. Another possibility is a change in the energy balance on small spatial scales resulting from a change in the local heating or cooling rates driven by local changes in the ionization. Cooling by the emission of optical and ultraviolet photons following electron collisional excitation of atoms and ions is proportional to the product  $n_en_{HI}$ \citep{Draine2011}. If the dominant heating process is from photoelectrons emitted by dust grains following absorption of UV photons, then the heating rate is proportional to the dust density and thus proportional to $n_{HI}$. If these two processes dominate the energy balance, then $n_e$ plays a critical role. Higher electron densities lead to enhanced cooling and thus lower temperatures, while lower electron densities lead to decreased cooling and thus higher temperatures. Support for this prediction comes from the \cite{Gry2001} analysis of the $\epsilon$~CMa sight line where $n_e$ is lower in the Blue cloud than the LIC but the temperature of the Blue cloud is higher.

All previous analyses of absorption lines in the LISM and the present work have assumed that the component of line broadening that depends on mass is thermal, so that line widths measure gas temperatures. Studies of the heliosphere indicate that small scale non-thermal processes such as shocks and charge exchange produce pickup ions that create a plasma with both thermal and supra-thermal velocities. Supra-thermal velocities are most easily detected in lines of low mass atoms such as D~I ($m$=2), because turbulent broadening dominates over thermal broadening for high mass ions such as Mg~II ($m$=24). Thus what we have called temperature may in fact be a combination of thermal and supra-thermal velocities, and the observed inhomogeneous ``temperatures" may be due in part to spatial differences in the supra-thermal broadening. Also, the different time scales for ionization and recombination for warm gas in the ISM cause the plasma to be out of equilibrium such that the electron temperature responsible for line broadening can be very different from a ``temperature" characterizing ionization balance \citep[e.g.,][]{deAvillez2012}. 

Small scale variations in the thermal and supra-thermal velocity structure can be maintained for a long time by inhomogeneous magnetic fields, since the Larmor radius of a 7,000~K thermal electron in a $3\mu$G magnetic field is 0.2~km and for a thermal proton is 350~km. Thus inhomogeneous structures smaller than 4,000~au can be maintained by random magnetic fields much smaller than $3\mu$G even for protons with energies much larger than thermal.

\section{Conclusions}

With the increase in the number of sight lines with analyzed interstellar properties, we have explored the diversity of these parameters within the LIC and other clouds in the CLIC with the following results:

\begin{enumerate}

\item Temperatures and turbulent velocities within the LIC and other nearby warm clouds have a wide range of values. The distributions appear to be random and can be fit with Gaussians. Within the LIC, temperatures range between 2,450~K and 12,900~K and turbulent velocities range between 0.0 and 4.4~km~s$^{-1}$. Since we measure average quantities along a sight line, the temperature and turbulent velocity measurements could under-represent the true inhomogeneities. The presence of temperature inhomogeneities implies density inhomogeneties as well. Previous methods of characterizing the properties of the LIC and other clouds by mean temperatures and turbulent velocities are not realistic and likely conceal important physical processes producing the inhomogeneity.

\item Comparison of temperature differences between pairs of sight lines across the LIC indicate that the angular scale for inhomogeneous temperatures is less than $2^{\circ}.2$. This angular scale corresponds to a linear scale of $\leq5,100$~pc, a distance that the Sun will traverse in less than 1000 years. The size and shape of the heliosphere will change when the Sun encounters changes in the interstellar medium.

\item Temperatures and turbulent velocities do not show any statistically significant trends with respect to the distance to the background star, Galactic coordinates, hydrogen column density, or angle relative to the direction of interstellar gas flowing into the heliosphere. The absence of a trend with stellar distance allows one to compare the properties of clouds in sight lines to stars located at a wide range of distances. The absence of trends with angle from the center of the LIC means that the properties of the center and the edge of the LIC are similar. The absence of trends in the temperature and turbulent velocity with angle relative to the strong EUV radiation source $\epsilon$~CMa indicates that photoionization by EUV radiation does not control the relative heating rate within the LIC or other clouds.

\item For 34 of the 37 sight lines to stars within 10~pc of the Sun, there is a unique match of each velocity component and direction with a previously identified warm cloud in the CLIC. If the density of neutral hydrogen is the same in all CLIC clouds as in the LIC near the Sun, $n_{HI}=0.20$~cm$^{-3}$, then none of these sight lines is completely filled with the identified clouds, and the clouds are separated by a fully ionized inter-cloud medium. If instead, $n_{HI}\approx0.10$~cm$^{-3}$ is typical for nearby clouds, then all of the sight lines within 4~pc are completely filled with warm clouds without separation by an ionized inter-cloud medium, but at larger distances the clouds are more widely separated. The question of the distance that the CLIC clouds extend into space from the Sun may be partially answered by a model in which the warm clouds completely fill space out to 4~pc and then become more widely separated by an ionized inter-cloud medium with distance to at least 10~pc.

\item A critical test of the neutral hydrogen density of the LIC is provided by the four stars within 4~pc of the Sun with sight lines that have only LIC absorption and astrospheres  that have Lyman-$\alpha$ absorption requiring that partially ionized hydrogen surrounds these stars. The most sensible model for these sight lines is that LIC gas complete fills these sight lines. The mean fill factor for these sight lines is 0.47, indicating that the mean $n$(H~I)=0.094~cm$^{-3}$. We conclude that the mean neutral hydrogen number density in the LIC and perhaps other warm clouds in the CLIC is about 0.10~cm$^{-3}$, and that the higher density of 0.20~cm$^{-3}$ surrounding the heliosphere is anomalous perhaps because of overlap with G cloud gas as proposed by Swaczyna et al. 2022).

\item Since warm gas in the nearby clouds is about one-third ionized, the clouds may not be described by steady-state theoretical models of the WNM and WIM, which assume either completely neutral or highly ionized gas with different heating and cooling processes. Both models predict maximum temperatures of about 10,000~K controlled by the rapidly rising cooling of Lyman-$\alpha$ emission with increasing temperature. We find a number of sight lines with $T>10,000$~K, but none with temperatures in excess of about 13,000~K. We also find no sight lines with temperatures definitely below 3,000~K given measurement errors. This is consistent with the predictions of steady-state theoretical models that the temperature regime between 3,000~K and 300~K is unstable at constant pressure.

\item We find 10 sight lines with relatively low temperatures, $T\leq 3,000$~K. These cool temperature sight lines traverse seven known clouds and two directions where no clouds are presently known. Three of the sight lines are in similar directions, suggesting that the low temperatures have a common origin. Since two of these sight lines do not traverse known clouds and the direction of the third may be outside of the LIC, we suggest that the regions of cool gas could be examples of CNM.

\item We list several possible processes that could be responsible for the local variations in temperature within the LIC and other partially ionized clouds. Heating by shocks is an obvious candidate, but we consider two other possibilities. One possibility is a model in which the dominant heating process in the partially ionized clouds is from energetic photoelectrons emitted by dust grains following the absorption of stellar UV photons, and the dominant cooling process is by electron excitation of atoms and ions followed by emission of optical and UV photons. The heating rate is proportional to $n_{HI}$ and the cooling rate is proportional to the product $n_en_{HI}$. In this case, the electron density and thus the relative ionization plays a critical role as increases in $n_e$ produce more cooling thereby lowering the temperature. Another possibility is that what is measured as ``temperature" may be a measure of both thermal and supra-thermal velocities with the non-thermal component produced by charge-exchange and other processes. The diversity in the measured ``temperatures" could be produced in part or entirely by spatial variations in the supra-thermal component. 

\end{enumerate}

\begin{acknowledgements}
We acknowledge support from the NASA Outer Heliosphere Guest Investigators Program to Wesleyan University and the University of Colorado for grant 80NSSC20K0785. The spectra used in this paper were obtained by {\em HST} primarily from the SNAP programs 10236, 11568, 13332, 13658, 14084, and 16487 but also from programs 13650 and 15071. The data were extract from MAST. We thank Brian E. Wood for his comments on the AD Leo data.

\end{acknowledgements}

Facilities: {\em HST}(STIS), {\em HST}(COS), {\em Voyager} I, {\em Voyager} II

\end{document}